\documentclass[conference]{IEEEtran}
\IEEEoverridecommandlockouts
\usepackage{cite}
\usepackage{amsmath,amssymb,amsfonts}
\usepackage{graphicx}
\usepackage{textcomp}
\usepackage{xcolor}
\usepackage{cite}
\usepackage{amsmath,amssymb,amsfonts}
\usepackage{graphicx}
\usepackage{textcomp}
\usepackage{xcolor}
\usepackage[hyphens]{url}
\usepackage{fancyhdr}
\usepackage{enumitem}
\usepackage[]{hyperref}
\usepackage{adjustbox}
\usepackage{graphicx}
\usepackage{subcaption}
\usepackage{caption}
\usepackage{float} % 引入 float 包以使用 [H] 选项
\usepackage{longtable}
\usepackage{multirow}
\usepackage{amsmath} % 包含文本圈的命令
\usepackage{tikz}
\usepackage{titlesec}
\usepackage{listings}
\usepackage{algorithm}
\usepackage[noend]{algpseudocode}  % 可去掉 noend，如果你想显示 EndIf 等
\usepackage{stfloats}
\usepackage{soulutf8}  % 比 soul 更稳定，支持 UTF-8
\soulregister\cite7    % 注册 \cite 命令，让 \hl 可以安全使用它
\soulregister\sysname7
\soulregister\autoref7
\sethlcolor{yellow} % 设置高亮颜色（可换成 cyan, lime, pink 等）
\usepackage{xcolor}  % 导言区中引入
\usepackage{booktabs}
\usepackage{marginnote}
\usepackage[normalem]{ulem}

\def\BibTeX{{\rm B\kern-.05em{\sc i\kern-.025em b}\kern-.08em
    T\kern-.1667em\lower.7ex\hbox{E}\kern-.125emX}}
\colorlet{softblue}{blue!20}
\colorlet{softorange}{orange!25}

\usepackage{titlesec}
\titlespacing*{\section}
{0pt}{4pt}{2pt}
\titlespacing*{\subsection}
{0pt}{3pt}{1pt}
\titlespacing*{\subsubsection}
{0pt}{2pt}{1pt}
\begin{document}

\title{\LARGE \bf 
% One Pool, Two Caches: Adaptive HBM Partitioning for Accelerating Generative Recommender Serving
When KV Meets Embeddings: Dynamic GPU Memory Allocation for Accelerating Generative Recommender Serving
% \\
% {\footnotesize \textsuperscript{*}Note: Sub-titles are not captured for https://ieeexplore.ieee.org  and

}
% \author{\normalsize{SC 2026 Submission
%     \textbf{\#128} -- Confidential Draft -- Do NOT Distribute!!}}

\author{
\IEEEauthorblockN{Wenjun Yu}
\IEEEauthorblockA{\textit{Hong Kong Baptist University} \\
Hong Kong, China \\
cswjyu@comp.hkbu.edu.hk}
\and
\IEEEauthorblockN{Shuguang Han}
\IEEEauthorblockA{\textit{Alibaba Inc} \\
Hangzhou, China \\
shuguang.sh@alibaba-inc.com}
\and
\IEEEauthorblockN{Amelie Chi Zhou\textsuperscript{*}}
\IEEEauthorblockA{\textit{Hong Kong Baptist University} \\
Hong Kong, China \\
amelieczhou@comp.hkbu.edu.hk}
}

\maketitle

\thispagestyle{fancy}
\lhead{}
\rhead{}
\chead{}
\lfoot{\footnotesize{
SC26, November 15-20, 2026, Chicago, Illinois, USA
\newline 979-8-3195-4789-7/26/\$31.00 \copyright 2026 IEEE}}
\rfoot{}
\cfoot{}
\renewcommand{\headrulewidth}{0pt}
\renewcommand{\footrulewidth}{0pt}

\begingroup
\renewcommand\thefootnote{} % 告诉 LaTeX 这个脚注“没有”编号
\footnotetext{
  \rule{2cm}{0.4pt}\vspace{0.4ex}
  
  \textsuperscript{*}Corresponding author.
}
\endgroup

\pagestyle{plain} % 打开页码

\newcommand{\sysname}{\emph{RACER}}
\newcommand{\allocator}{\emph{SmartAllocator}}

% \pdfpagewidth=8.5in
% \pdfpageheight=11in
% \setlength{\textfloatsep}{1pt plus 1.0pt minus 1.0pt}
% % \setlength{\floatsep}{1pt plus 1.0pt minus 1.0pt}
% \setlength{\intextsep}{1pt plus 1pt minus 1pt} 

\begin{abstract}
\label{re:abstract}
Generative Recommender (GR) inference faces a dual memory bottleneck absent in LLM serving. Embedding hot caches and KV caches compete for limited GPU HBM, while existing systems optimize them in isolation, leading to suboptimal tail latency. We observe that (1) the optimal EMB--KV memory allocation ratio shifts by up to 0.35 across workload regimes, and static allocation leaves 20–30\% latency improvement untapped, while continuous memory reallocation risks cache thrashing and P99 SLO violations. We present \sysname{}, a system that jointly and adaptively reallocates HBM between EMB and KV caches at runtime. \sysname{} addresses two core challenges: (1) \textit{Adaptive Memory Allocation} via a three-layer PPO controller with online adaptation and burst recovery, achieving 32\,$\mu$s decision latency; and (2) \textit{Embedding-KV-Aware Scheduling} that routes requests to maximize joint EMB--KV cache reuse across serving nodes. 
Evaluations on three production-scale datasets show that \sysname{} reduces P99 latency by 24–38\% compared to the best static allocation strategy. It also achieves 93.5–99.6\% SLO satisfaction across workload regimes, significantly outperforming state-of-the-art baselines (67–94.2\%) without sacrificing serving throughput. 
\end{abstract}

% \begin{IEEEkeywords}
% Generative Recommender,
% Large-Scale Inference Serving,
% GPU Memory Management,
% Resource Scheduling,
% Performance Optimization
% \end{IEEEkeywords}

% \vspace{-12pt}
\section{Introduction}
\label{sec:introduction}

%gr越来越重要 例如hstu被很多厂商采用，而且精度远超传统方法，hstu类的gr模型主要由两部分组成
Generative Recommenders (GRs) are transformer-based recommendation models that predict the next item or a set of candidate items from a user’s historical interaction sequence~\cite{zhai2024actions,zhou2025onerec}.
In contrast to traditional recommendation models such as DLRMs~\cite{naumov2019deep}, which primarily rely on shallow feature interactions, GRs leverage transformer-based sequence modeling over long user interaction histories (10k+), achieving higher accuracy~\cite{guan2025make,chai2025longer,zhou2026gems}.
GRs have become increasingly important in large-scale production recommendation systems. In this paper, we focus on Meta's HSTU~\cite{zhai2024actions} as a representative GR architecture due to its widespread industrial adoption~\cite{liang2025tbgrecall,huang2025towards,han2025mtgr}, making our insights broadly applicable. %It is one of the most popular GR models widely adopted in industry (e.g., Alibaba~\cite{liang2025tbgrecall}, Xiaohongshu~\cite{huang2025towards}, and Meituan~\cite{han2025mtgr}).

HSTU inference consists of two tightly coupled stages. \emph{1) Embedding Table Lookup} retrieves dense vectors for sparse features from large embedding tables sharded in CPU DRAM, whose scale reaches tens to hundreds of terabytes (e.g., 96TB Model-F and 200TB+ Persia~\cite{mudigere2022software,lian2022persia}), leading to frequent DRAM–HBM transfers and memory pressure. \emph{2) Transformer Block Computation} applies stacked multi-head self-attention
over long interaction sequences, with KV cache reuse retained in HBM to avoid
redundant recomputation.

% Production GR架构
Production GR serving clusters comprise a \textit{request router} and multiple \textit{serving nodes} (see \autoref{fig_hstu_serving}), each featuring a two-level memory hierarchy with GPU HBM and CPU DRAM~\cite{sun2025xgr,kim2025characterization,sun2026bat}. Within HBM, two persistent caches co-exist: \emph{1) embedding hot caches} that store frequently accessed embeddings (EMB) to reduce PCIe/RDMA transfers from CPU-side sharded EMB tables, and \emph{2) KV caches} that retain precomputed attention tensors for historical sequences, avoiding redundant recomputation. As a result, both EMB and KV caches are inside the same GPU HBM memory.

\begin{figure}[t]
  \centering
  \begin{subfigure}[t]{0.49\linewidth}
    \centering
    \includegraphics[width=0.9\linewidth]{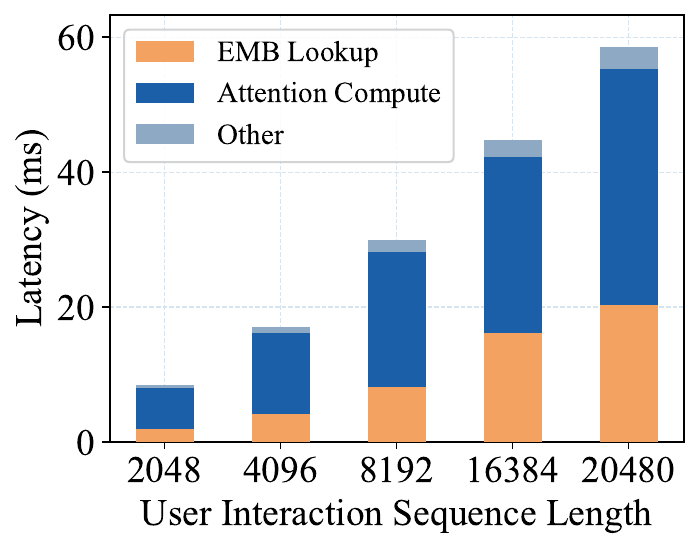}
    \caption{Latency breakdown with various sequence length (\#table=10)}
    \label{fig_emb_length_breakdown}
  \end{subfigure}
  \hfill
  \begin{subfigure}[t]{0.49\linewidth}
    \centering
    \includegraphics[width=0.9\linewidth]{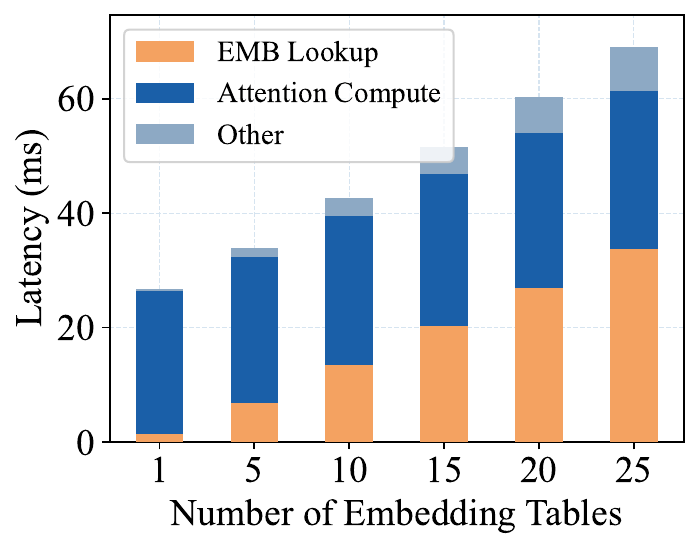}
    \caption{Latency breakdown with various tables (\#seq. length=15k)}
    \label{fig_table_breakdown}
  \end{subfigure}
  \caption{HSTU inference latency under different settings}
  \label{fig_hstu_breakdown}
\end{figure}

% 计算和embedding lookup都是bottleneck
Profiling from production traces on real-world datasets reveals that both EMB lookup and KV cache access are primary bottlenecks, as shown in~\autoref{fig_hstu_breakdown}. EMB lookup consistently dominates end-to-end latency as interaction length scales to 10k+~\cite{guan2025make,chai2025longer,zhou2026gems}: at a sequence length of 20,480, it alone contributes over 40\% of total inference time. The KV computation phase also incurs substantial overhead due to long-sequence attention at the same time. The figure further shows that each additional embedding table introduces a roughly linear transfer cost, since every item must be looked up across all tables and production deployments commonly involve tens to hundreds of tables~\cite{avazu,kdd12,criteo1tb_click_logs, Zha2022AutoShard}.

% 现如今方法只优化了单方面
Existing systems address each bottleneck in isolation. On the embedding side, prior work focuses on prefetching~\cite{zhao2023recd}, sharding~\cite{ren2025machine}, and near-storage access~\cite{yu2026near} to reduce EMB lookup latency. On the KV side, recent GR-aware systems propose KV compression, sharing, and cache-aware routing~\cite{sun2025xgr,kim2025characterization,sun2026bat} to improve reuse rates. However, both lines of work implicitly treat HBM capacity as fixed for their respective caches, ignoring the fundamental resource contention between them. Allocating more HBM to the embedding hot cache reduces EMB lookup stalls but starves the KV cache, increasing recomputation overhead and vice versa. No existing system jointly manages this trade-off, leaving significant latency reduction unrealized. 

% 我们发现hbm只给一方不行 所以需要动态优化
% 需要先说我们的motivation
Insights from production metrics indicate that allocating all HBM to a single cache is suboptimal. As shown in \autoref{fig_emb_kv_partition_all}, the optimal EMB–KV HBM memory allocation is highly dynamic across sequence lengths and workload scenarios, with the ratio shifting from ${\approx}0.20$ to ${\approx}0.55$ as sequence length increases or workloads vary. This leaves 20–30\% latency improvement untapped under static allocations. 
The optimal memory allocation evolves over time, and adapting it online can interfere with live inference. Memory reallocation may evict hot embeddings or starve in-flight prefill requests, leading to latency spikes that violate strict production P99 SLOs (30–50 ms~\cite{wang2026relaygr,yang2025gpu}).

% 我们发现hbm只给一方不行 所以需要动态优化
% 然后再说遇到什么问题 我们怎么设计的

These observations surface three concrete challenges that our system must
address. 
\emph{(1) Multi-factor workload modeling:} The optimal memory allocation ratio is jointly determined by the hot-user ratio, KV hit rate, and EMB hit rate, which together account for over 60\% of decision importance, making simple heuristics insufficient.
\emph{(2) Fast and non-intrusive adjustment:} even a 10\% ratio change can trigger up to 8GB H2D data movement on the critical path of embedding transfers, degrading service quality and requiring incremental, decoupled allocation updates. 
\emph{(3) Joint-aware scheduling:} once memory allocation ratios become heterogeneous across nodes, EMB transfer costs vary per node, breaking KV-only routing and requiring EMB--KV–aware scheduling.

% 我们的贡献
We present \sysname{} 
(\textbf{R}einforcement-based \textbf{A}daptive 
\textbf{C}ache allocation and routing \textbf{E}ngine for 
\textbf{R}ecommendation)
% (\textbf{H}BM-level \textbf{E}MB–KV \textbf{L}atency \textbf{M}anager)
, which jointly allocates GPU HBM between EMB and KV caches at runtime and routes requests to maximize cache reuse, while satisfying strict production P99 SLO targets (30–50 ms) without disrupting live inference traffic. 
On three production-scale datasets over a 32-node A100 cluster (5TB EMB tables, 8K–15K sequence lengths), \sysname{} reduces P99 latency by 24–38\% over the best static allocation, achieves ${\geq}93.5\%$ SLO satisfaction, and incurs only 32~$\mu$s overhead. 
We summarize our key contributions as follows:
\begin{itemize}[leftmargin=*, topsep=2pt, itemsep=2pt, parsep=0pt]
    \item \textbf{Dual-bottleneck characterization for GR serving.}
    We identify the EMB--KV memory allocation bottleneck in GR inference and show that the optimal memory allocation ratio $\alpha^*$ shifts by up to 0.35, leaving 20--30\% latency improvement untapped under static policies.
    \item \textbf{Adaptive EMB--KV memory allocation.}
    We design a lightweight runtime controller that tracks the best EMB--KV memory allocation under dynamic workloads, achieving 32\,$\mu$s decision latency and staying within 0.024--0.029 of the offline-optimal ratio across evaluated settings.
    \item \textbf{Non-disruptive memory allocation adjustment.}
    We present a memory-management mechanism that applies allocation updates under live traffic without relocating resident KV state or introducing uncontrolled H2D interference. 
    \item \textbf{EMB--KV-aware request routing.}
    We design a routing policy that jointly considers KV locality, embedding locality, and node load, yielding an additional 15--20\% latency reduction under burst traffic. 
\end{itemize}

% \vspace*{-1ex}
\section{Background and Motivation}
\label{sec:background}  % 注意小写一致

\subsection{Generative Recommender Systems}
\begin{figure}[t]
  \centering
  \includegraphics[width=0.75\linewidth]{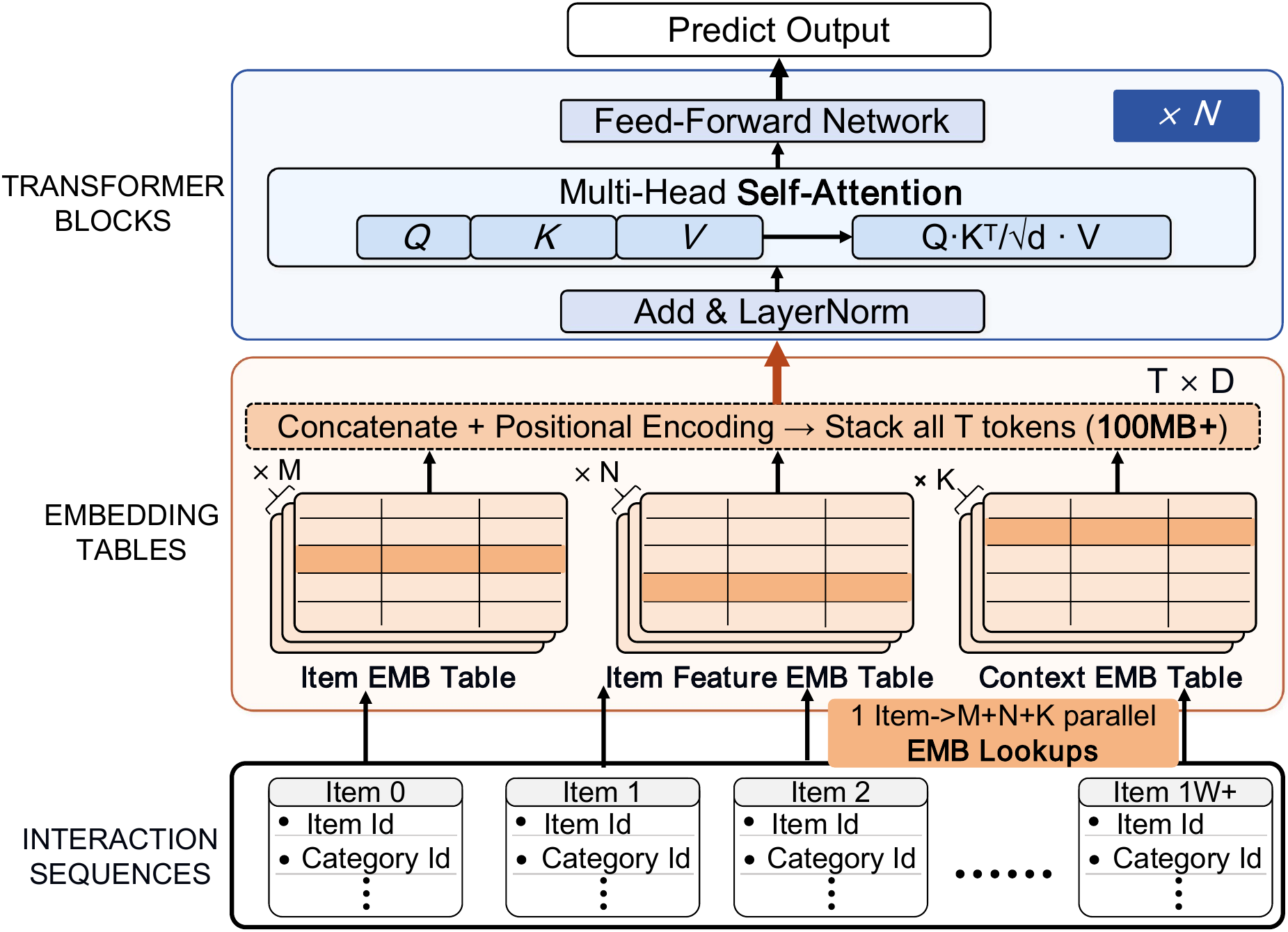}
  \caption{HSTU Model Structure}
  \label{fig_hstu_structure}
\end{figure}

Generative Recommenders (GRs) are transformer-based recommendation models that predict future user interactions from long historical behavior sequences. In this paper, we focus on Meta’s HSTU~\cite{zhai2024actions} as a representative GR architecture because it has become a common design point in industrial generative recommendation systems. As illustrated in \autoref{fig_hstu_structure}, HSTU takes an ordered sequence of prior \textbf{user-item interactions} as input and produces candidate items as recommendation output. %by combining embedding lookup with stacked self-attention blocks.

HSTU inference has two stages that place different demands on the memory system. 
\begin{itemize}[leftmargin=*]
    \item First, the model performs \textbf{embedding lookups} for each item and its associated categorical features (e.g., item ID, category, and context). These embeddings reside in large embedding tables that can reach tens to hundreds of terabytes in production~\cite{coburn2025meta,dlrmv3_2026, mudigere2022software,lian2022persia}, far exceeding GPU memory capacity. Consequently, the tables are typically sharded across CPU memory, and cache misses require data movement from host DRAM or remote nodes. 
    \item Second, the retrieved embeddings are processed by stacked \textbf{transformer blocks}, whose self-attention layers operate over long interaction histories (e.g., over 10K+ items in modern GR systems) and therefore incur substantial compute and memory cost.
\end{itemize}

A key property of GR inference is that both stages benefit from persistent GPU-resident state. Frequently accessed embedding vectors can be retained in an embedding cache (EMB cache) to reduce repeated DRAM-to-HBM transfers. Likewise, previously computed key/value tensors can be retained in a KV cache to avoid recomputing attention state for users whose histories overlap with earlier requests. These two caches accelerate different parts of the pipeline, but they compete for the same limited GPU HBM capacity. 
This competition distinguishes GR serving from prior LLM/DLRM systems that optimize only one cache in isolation~\cite{sun2026bat,ren2025machine}.

\subsection{Online Serving Architecture of GRs}
% 现有工作为什么在gr kv hit rate上面不work 要介绍prompt不一样
% 要画一个现有的hstu serving系统 要画出来prompt是什么样 cpu-gpu
% serving pipeline是什么样
\begin{figure*}[t]
  \centering
  \includegraphics[width=0.8\linewidth]{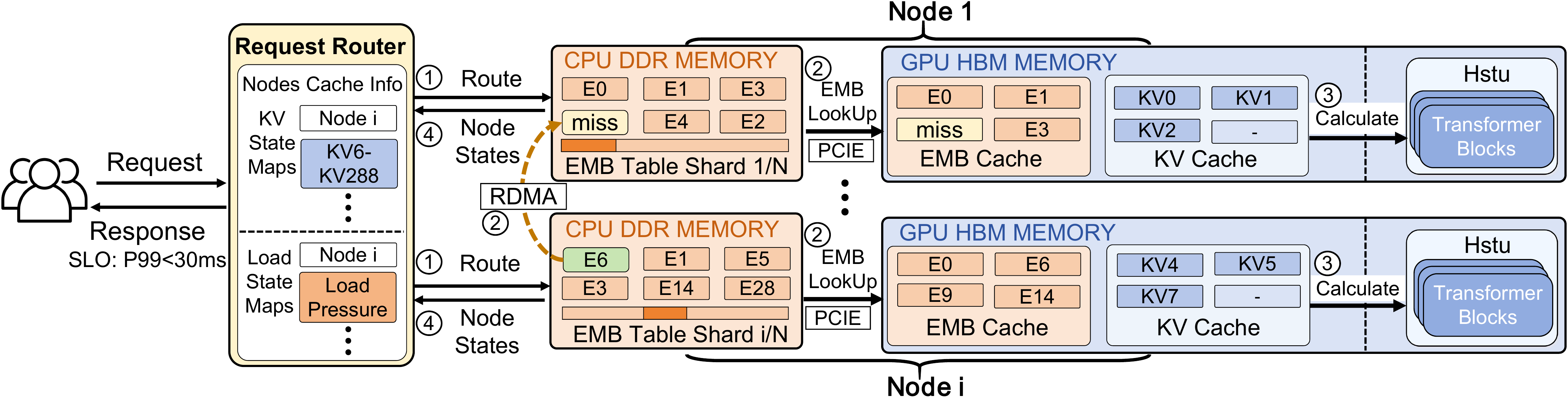}
  \caption{HSTU Online Inference Serving Architecture}
  \label{fig_hstu_serving}
\end{figure*}

Production GR systems typically adopt a multi-node serving architecture to satisfy strict tail-latency service-level objectives (SLOs), commonly $\sim$30--50\,ms in industry deployments~\cite{wang2026relaygr,yang2025gpu}. As shown in \autoref{fig_hstu_serving}, a typical GR serving deployment consists of two core components~\cite{sun2025xgr,kim2025characterization,sun2026bat}, namely a \textbf{request router} and a set of \textbf{serving nodes}. Each node contains CPUs for orchestration and host-side memory management, together with GPUs for embedding access and transformer execution. This organization creates a two-level memory hierarchy spanning CPU DRAM and GPU HBM across the cluster. 

\textbf{Data placement.}
Within this memory hierarchy, data is naturally split by capacity and access pattern. Because embedding tables are far larger than a single node’s memory capacity, they are \emph{sharded} across CPU DRAM over all serving nodes. In contrast, each node’s GPU HBM holds two persistent caches: an {EMB cache} for hot embeddings, and a {KV cache} for reusable attention state. 
Unlike LLM serving, where KV movement can often be overlapped with autoregressive decoding, GR inference in our setting is executed as a single forward pass with no comparable decode phase. As a result, CPU offloading of KV states~\cite{Gao2024Fast,pan2025instattention} is unattractive for our workloads, and KV states are stored in GPU only during serving.

\textbf{Request serving.}
Under this architecture, each request proceeds through four stages: \emph{(1) Route:} the request router selects a serving node; \emph{(2) EMB lookup:} if the required embeddings are absent from that node’s EMB cache, they are fetched from the local CPU shard or from a remote node through RDMA; \emph{(3) Compute:} the GPU executes the HSTU forward pass, reusing any resident KV state and cached embeddings; and \emph{(4) State sync:} after completion, the serving node asynchronously reports updated cache residency and load information back to the router. This request path makes both embedding misses and KV reuse directly visible in end-to-end latency.

% This serving architecture exposes the key systems tension studied in this paper. Since both embedding hot data and KV states occupy GPU HBM, allocating more memory to one cache directly reduces the capacity of the other.
% To quantify how this tension manifests in practice, we profile a 6-layer HSTU model with 512-dimensional embeddings on our production traces, while varying sequence length and embedding-table count. To isolate the embedding-side cost from KV reuse, we disable KV cache reuse during profiling. 
% As shown in \autoref{fig_emb_length_breakdown}, EMB lookup remains a major component of end-to-end latency across settings and becomes the dominant bottleneck at long sequence lengths: at 20{,}480 tokens, it contributes over 40\% of total latency. \autoref{fig_table_breakdown} shows a similar trend: the lookup latency grows almost linearly with the EMB table count and production deployments commonly involve tens to hundreds of tables~\cite{avazu,kdd12,criteo1tb_click_logs,Zha2022AutoShard}. 

% These results show that GR serving cannot be effectively optimized by improving KV reuse alone~\cite{sun2025xgr,kim2025characterization,sun2026bat}: even when KV-side techniques reduce attention recomputation, embedding access can still emerge as the dominant bottleneck in important workload regimes. More fundamentally, because EMB and KV contend for the same HBM budget, optimizing either side in isolation can simply shift the bottleneck to the other rather than improve end-to-end performance, motivating a joint approach to memory management.

\begin{figure}[t]
  \centering
  \begin{subfigure}[t]{0.49\linewidth}
    \centering
    \includegraphics[width=\linewidth]{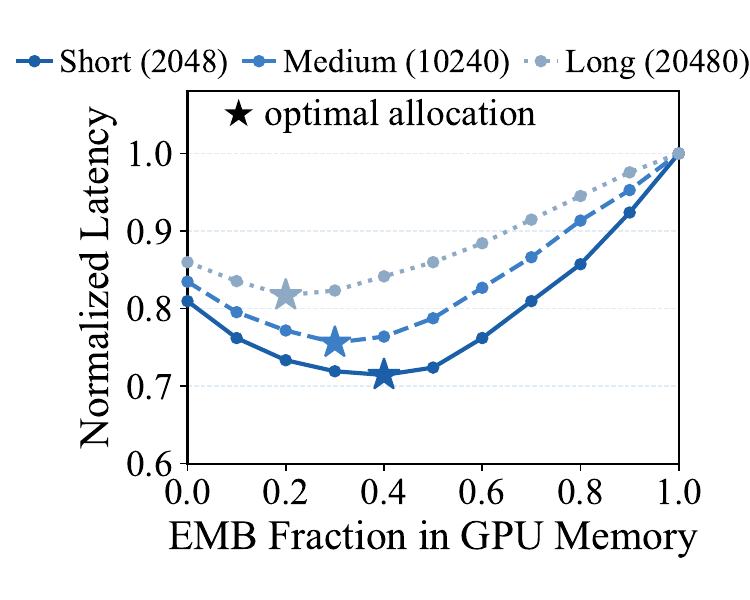}
    \caption{Varying sequence lengths (same scenario)}
    \label{fig_emb_kv_best_partition}
  \end{subfigure}
  \hfill
  \begin{subfigure}[t]{0.49\linewidth}
    \centering
    \includegraphics[width=\linewidth]{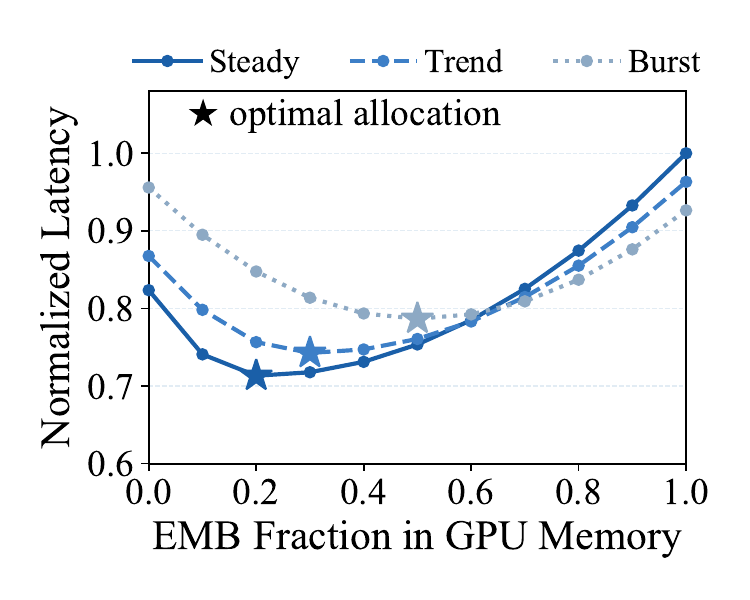}
    \caption{Varying scenarios (fixed sequence length)}
    \label{fig_optimal_change}
  \end{subfigure}
  \caption{Normalized latency vs.\ EMB--KV allocation ratio.}
  \label{fig_emb_kv_partition_all}
\end{figure}

\subsection{Motivation and Design Challenges}
\label{sec_motivation_challengs}

The serving architecture of GRs exposes the key system tension studied in this paper: the \textbf{EMB hot cache}, which reduces the EMB fetch cost, and the \textbf{KV cache}, which avoids redundant attention recomputation, compete for limited GPU HBM resources.
Since both caches contribute to the end-to-end serving latency, the key system question is not how to optimize either cache in isolation, but how to divide limited HBM capacity between them and manage their trade-off online.

To answer this question, we empirically study the cost of the two caches both individually and jointly.

\textbf{First}, to isolate EMB overhead from KV reuse, we profile a 6-layer HSTU model with 512-dimensional embeddings while varying sequence length and EMB-table count, with KV cache reuse disabled during this measurement.~\autoref{fig_hstu_breakdown} shows that the EMB lookup remains a major component of end-to-end latency across settings and becomes dominant for long sequences and large tables. For example, with a sequence length of 15k and 25 embedding tables, the EMB lookup alone accounts for more than 50\% of total latency. Importantly, the figure shows that the EMB lookup cost grows roughly linearly with the number of embedding tables, making this a severe issue in production deployments that commonly involve tens to hundreds of tables~\cite{criteo1tb_click_logs,Zha2022AutoShard}.

\textbf{Second}, we quantify the joint trade-off between the two caches by sweeping the fraction of GPU HBM allocated to embeddings.~\autoref{fig_emb_kv_partition_all} shows that end-to-end latency is minimized at an interior allocation rather than at either extreme: allocating too little HBM to the EMB cache increases embedding misses and transfer stalls, whereas allocating too little HBM to the KV cache reduces reusable history and increases recomputation. More importantly, the optimal allocation shifts substantially across workloads, moving from approximately 0.20 to 0.55 as sequence length and workload conditions change. This behavior implies that no single static allocation can remain near-optimal across operating regimes, leaving up to 20–30\% latency improvement untapped in our study.

% These results show that GR serving cannot be effectively optimized by improving KV reuse alone~\cite{sun2025xgr,kim2025characterization,sun2026bat}: even when KV-side techniques reduce attention recomputation, embedding access can still emerge as the dominant bottleneck in important workload regimes. More fundamentally, because EMB and KV contend for the same HBM budget, optimizing either side in isolation can simply shift the bottleneck to the other rather than improve end-to-end performance, motivating a joint approach to memory management.
\begin{figure}[t]
  \centering
  \begin{minipage}[t]{0.49\linewidth}
    \centering
    \includegraphics[width=\linewidth]{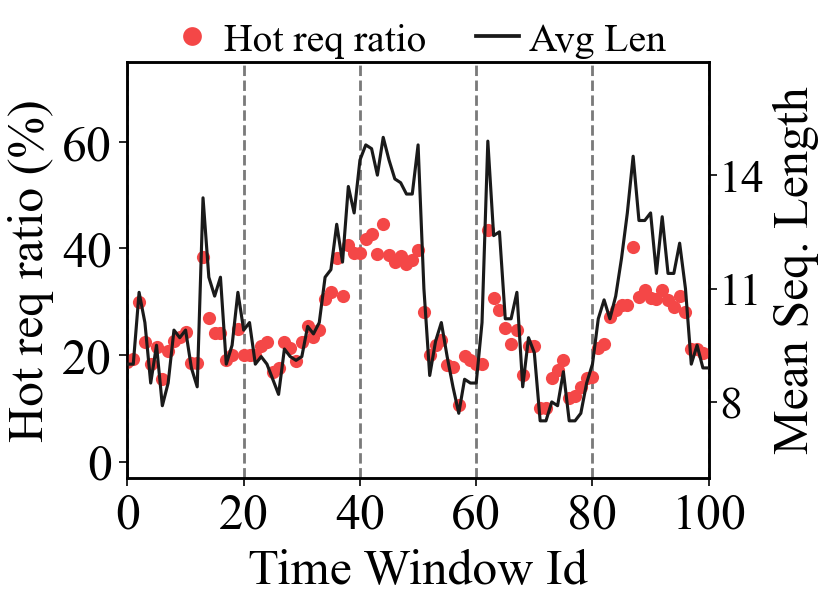}
    \caption{Rapid production fluctuations in hot-user req. share and mean seq. length}
    \label{fig_dynamic_workload}
  \end{minipage}
  \hfill
  \begin{minipage}[t]{0.49\linewidth}
    \centering
    \includegraphics[width=\linewidth]{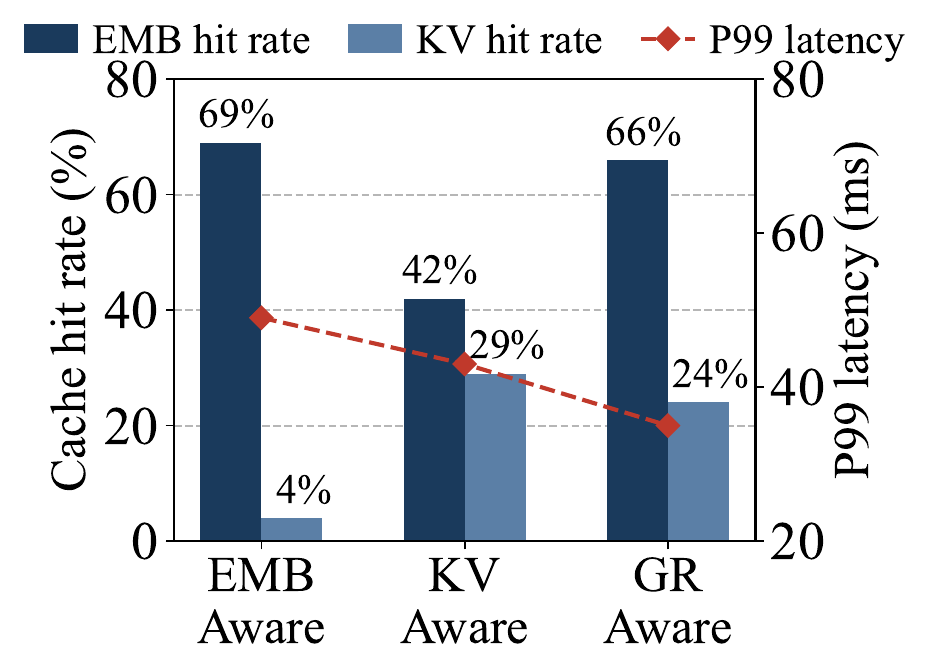}
    \caption{Comparison of cache hit rates across three scheduling methods}
    \label{fig_caches}
  \end{minipage}
\end{figure}

Together, these observations induce three design challenges:

\textbf{Challenge 1: The optimal EMB--KV allocation is dynamic, making static allocation insufficient.}
As shown in \autoref{fig_emb_kv_partition_all}, {the optimal EMB--KV memory allocation is not fixed}, and can be affected by many factors (e.g., user-item interaction sequence length, EMB-table count). Further, \autoref{fig_dynamic_workload} shows the temporal workload variation using our production trace. The window size is 5\,s. The figure reveals that the workload composition can change abruptly even across short windows, making memory reallocation a recurring need rather than a rare event.
This result implies that EMB--KV management must adapt \emph{quickly online} based on runtime workload signals rather than rely on a fixed allocation.

\textbf{Challenge 2: Online memory reallocation is necessary, but naive adjustment can disrupt live inference.}
Effective GR serving requires online HBM memory reallocation between EMB hot cache and KV cache.
However, reallocation under live traffic is not free. 
A naive memory allocation scheme that expands one side of HBM must evict or refill cache state on the other side, introducing background data movement on the same host-to-device (H2D) path used by latency-critical EMB lookups. In our profiling, burst periods can generate $\sim$10 GB/s of H2D traffic for embedding misses alone, so naive memory reallocation risks increasing queueing delay and jitter on the critical path. Under \emph{strict production P99 SLO targets}, such interference can be unacceptable even when average bandwidth utilization is not saturated.
%Therefore, the challenge is not merely to react to workload changes, but to adapt the EMB--KV memory allocation in a safe, incremental, and non-intrusive manner that preserves tail-latency stability.

%This challenge is further amplified by workload skew: as shown in \autoref{fig_hot_user}, hot users can account for a large fraction of requests (e.g., top-5\% users account for 38\% total traffic in Amazon Books~\cite{}), so shifts in hot-user concentration can rapidly change KV pressure and move the optimal allocation on short timescales. 
%In addition, repartitioning competes with the serving path itself: background refill or migration activity shares the same host-to-device path as latency-critical embedding fetches, especially during burst periods when bandwidth is already heavily utilized. Thus, the problem is not merely to react to workload changes, but to adapt the EMB--KV partition in a way that preserves tail-latency stability.

\textbf{Challenge 3: Memory allocation and request routing are inherently coupled.}
The EMB--KV trade-off is not purely a node-local problem. 
In production GR serving, EMB tables are sharded across nodes, while both hot embedding sets and KV states are distributed throughout the cluster. Request routing therefore determines where user-specific KV state and embedding locality accumulate, which in turn affects each node’s effective EMB and KV hit rates.
\autoref{fig_caches} illustrates this coupling.
An embedding-aware scheduler~\cite{li2025embedding} improves EMB hit rate but sacrifices KV locality, while a KV-aware scheduler~\cite{sun2025xgr} improves KV hit rate but ignores embedding locality. Our scheduler that jointly considers both achieves a more balanced hit-rate profile and, more importantly, lower end-to-end latency than either single-objective policy. These results show that memory allocation and request routing cannot be optimized independently.

These challenges highlight why isolated optimizations are insufficient. Embedding-side techniques such as prefetching~\cite{ren2025machine} and KV-side techniques such as KV sharing~\cite{sun2025xgr,sun2026bat} or compression~\cite{kim2025characterization} improve only one side, but do not eliminate contention between the EMB and KV caches for GPU HBM. Insufficient EMB allocation causes prefetched embeddings to be evicted early and prefetch traffic to compete with H2D transfers, while more KV allocation shrinks EMB capacity and increases misses. Optimizing either alone shifts the bottleneck, so effective GR serving requires joint runtime control of allocation, memory, and request routing.

\section{Design Rationale}\label{sec:rationale}

The challenges identified above suggest that efficient GR serving must address not one isolated bottleneck, but three coupled design problems: \textbf{1)} deciding the EMB–KV split online as workloads change, \textbf{2)} applying allocation changes without disrupting latency-sensitive inference, and \textbf{3)} routing requests using the resulting per-node memory conditions.
To address these problems, we propose \sysname{}, which is designed based on three rationales. 

% Figure~\ref{fig_our_system} illustrates the overall organization of our system.
% At a high level, it forms a closed runtime loop between online measurement, allocation control, safe memory actuation, and request routing: each serving node monitors recent cache and latency signals, a node-local allocator decides how GPU HBM should be divided between the EMB cache and KV cache, the memory manager applies that decision incrementally, and a global scheduler uses refreshed cache-state summaries to guide subsequent request placement. 

\uline{\textbf{Rationale 1}: Dynamic memory reallocation requires lightweight and predictive online control.}
% Dynamic EMB--KV memory allocation is fundamentally an online control problem. The controller must react quickly to workload shifts, but the quality of a memory allocation update is revealed only after cache refill, eviction, and future reuse alter hit rates and tail latency. 
% As a result, the controller must make lightweight decisions from several coupled runtime signals while accounting for delayed consequences rather than optimizing a single instantaneous metric.
% This combination of responsiveness, multi-factor dependence, and delayed feedback makes fixed heuristics and purely reactive policies insufficient. In this paper, we design a lightweight Reinforcement-Learning (RL) based controller, which can make fast online decisions while optimizing for the downstream effect of each memory reallocation update on serving performance. (Details in Section~\ref{subsec_partition_model})
Dynamic EMB--KV allocation is fundamentally an online control problem: the controller must react to workload shifts, while the impact of each update is only realized after cache refill, eviction, and future reuse affects hit rates and tail latency.
Thus, it must make lightweight decisions from coupled runtime signals while accounting for delayed effects, rather than optimizing a single instantaneous metric.
This combination makes fixed heuristics and purely reactive policies insufficient.
We therefore design a lightweight RL-based controller that enables fast online decisions while optimizing the downstream impact of each reallocation on serving performance (Section~\ref{subsec_partition_model}).

% \uline{\textbf{Rationale 2}: Memory reallocation adjustment should be fast and non-disruptive.}
% Dynamically adapting EMB--KV memory reallocation is useful only if it can be realized under live traffic without degrading serving quality. 
% In production GR serving, however, memory reallocation competes with latency-critical embedding misses for the same H2D transfer path, so a naive adjustment can worsen the very P99 latency it is intended to improve. 
% This makes memory reallocation an actuation problem rather than a one-time configuration step. A practical design must minimize disruptive data movement, decouple allocation updates from serving-critical transfers, and amortize adjustment cost across time instead of relying on bulk reconfiguration. \sysname{} is designed around these principles and uses a lightweight online memory-management mechanism to enact allocation changes incrementally and safely. (Details in Section~\ref{subsec_memory_management})

\uline{\textbf{Rationale 2}: Memory reallocation should be fast and non-disruptive.}
Dynamic EMB--KV reallocation is only useful if it can run under live traffic without harming serving quality.
In production, reallocation competes with latency-critical embedding misses for the same H2D path, so naive adjustments can worsen P99 latency.
This makes reallocation an actuation problem rather than a one-time configuration. A practical design must minimize disruptive data movement, decouple updates from critical transfers, and amortize adjustment cost over time.
\sysname{} follows these principles with a lightweight online memory-management mechanism that applies allocation changes incrementally and safely (Section~\ref{subsec_memory_management}).

\uline{\textbf{Rationale 3}: Request routing should capture the joint EMB--KV trade-off.}
In production GR serving, the router decides which serving node should handle each incoming request. This decision becomes more difficult under dynamic EMB--KV memory reallocation, because nodes no longer look identical due to embedding table sharding and distribution of EMB/KV caches. Request routing must therefore reason jointly about KV locality, embedding locality, and load balancing. This makes existing KV-centric schedulers insufficient. In this paper, \sysname{} uses an EMB--KV-aware scheduler that explicitly estimates request cost from both cache perspectives and routes requests according to expected end-to-end latency. (Details in Section~\ref{subsec_scheduling})

\begin{figure}[t]
  \centering
  \includegraphics[width=0.72\linewidth]{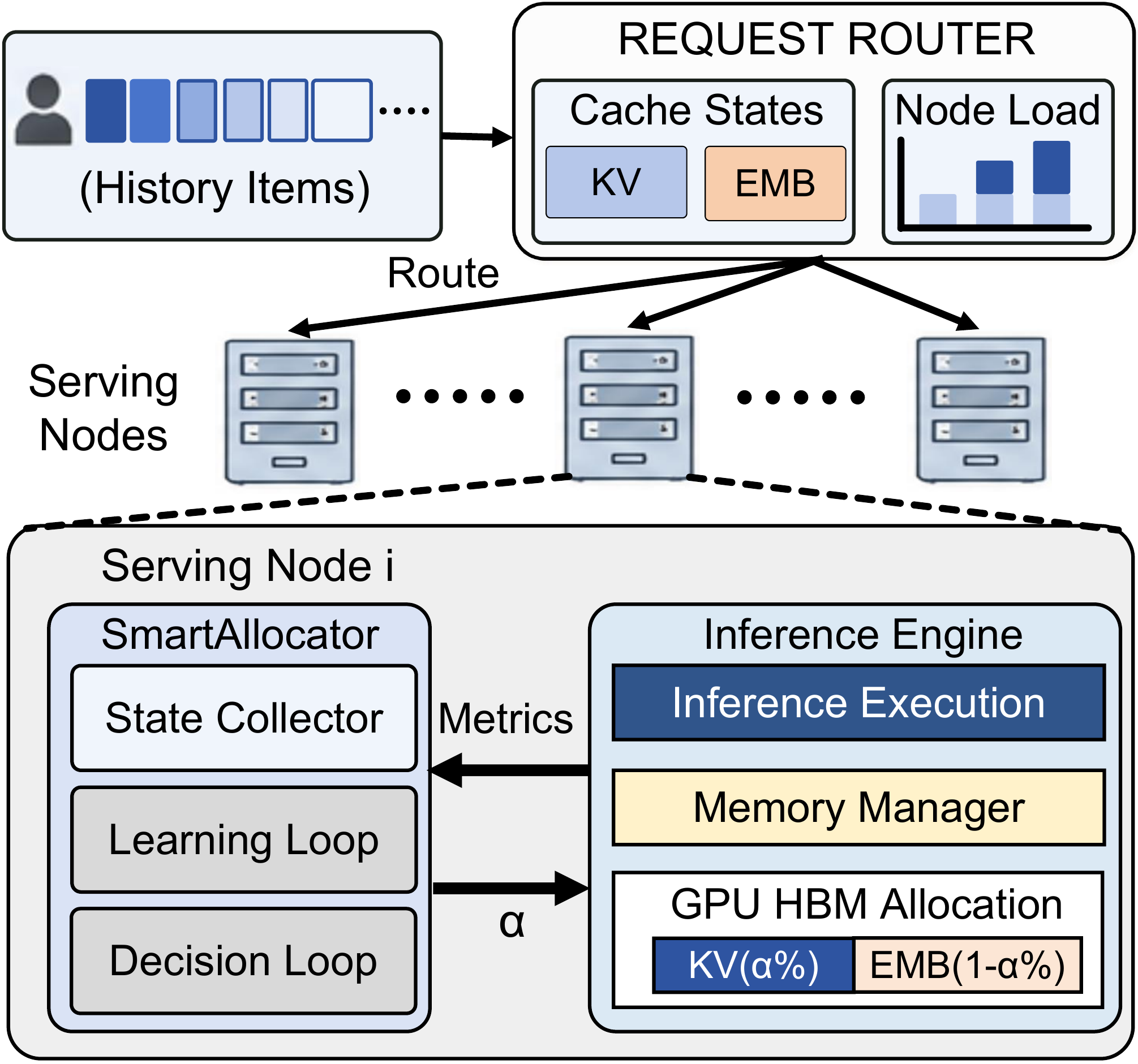}
  \caption{Overall architecture of \sysname{}}
  \label{fig_our_system}
\end{figure}

% As shown in~\autoref{fig_our_system}, \sysname{} is designed with two interacting roles: a \emph{global request router} and a \emph{node-local control path}. Each serving node continuously monitors recent workload, cache, and latency signals, and the node-local controller uses them to determine how GPU HBM should be divided between the EMB cache and the KV cache. 
% The resulting memory allocation update is then applied incrementally by the memory manager so that adjustment overhead remains decoupled from latency-critical inference traffic. At the cluster level, the global request router uses refreshed cache-state summaries together with node load to route subsequent requests according to their expected end-to-end serving cost. In this way, \sysname{} connects runtime measurement, allocation control, safe memory management, and routing into a single coordinated system for GR serving. 

The EMB–KV HBM allocation is not governed by a single component. As shown in~\autoref{fig_our_system}, \sysname{} forms a tightly entangled control loop between a \emph{global request router} and a \emph{node-local control path}. Each serving node continuously monitors workload, cache state, and latency signals, then the node-local \allocator{} uses these signals to dynamically determine how GPU HBM is partitioned between the EMB cache and the KV cache. These signals are in turn influenced by the router’s earlier placement decisions, which shape user locality and cache residency on each node. The resulting allocation ratio $\alpha$ is applied incrementally by the memory manager to avoid interfering with latency-critical inference. At the cluster level, the global request router leverages refreshed cache states, node load, and per-node $\alpha$ to estimate the end-to-end serving cost of each request, while accounting for cache heterogeneity as nodes may specialize differently under varying allocations. These routing decisions subsequently reshape cache state and runtime signals, feeding back into future allocation updates. In this way, \sysname{} connects runtime measurement, allocation control, safe memory management, and routing into a single coordinated system for GR serving.

\section{Design Details}
\label{design_details}

\subsection{Adaptive EMB--KV Memory Allocation}
\label{subsec_partition_model}

This section presents \allocator{}, a node-local adaptive memory allocator in \sysname{}.
At each decision epoch, \allocator{} determines how to partition GPU HBM between the EMB cache and KV cache based on recent runtime state. 
We formalize this as a sequential decision problem that minimizes end-to-end P99 latency under dynamic workloads.

\textbf{Problem Formulation.}
Denote the GPU memory budget as $M$ and the EMB allocation ratio as $\alpha \in [0,1]$. The memory capacity split for EMB cache and KV cache is $M_{\mathrm{Emb}}=\alpha\cdot M$ and $M_{\mathrm{KV}}=(1-\alpha)\cdot M$, respectively.
At each decision epoch $t$, our objective is to select the best $\alpha_t^*$ that minimizes end-to-end P99 latency $\ell(\alpha, s_t)$ under the current runtime state $s_t$:
\begin{equation}
\small
  \alpha_t^* = \arg\min_{\alpha \in [0,1]} \; \ell(\alpha,\, s_t).
  \label{eq:alloc_objective}
\end{equation}

\textbf{RL-based Solution Formulation.}
% To capture the online optimization process, we model the dynamic EMB--KV memory allocation as a Markov Decision Process (MDP), in which \allocator{} observes the current runtime state, selects an allocation update, and receives reward from the resulting serving performance in subsequent windows.
% Formally, we define the MDP as a tuple $(\mathcal{S}, \mathcal{A}, \mathcal{P}, \mathcal{R}, \gamma)$, where $\mathcal{S}$ is the state space, $\mathcal{A}$ is the action space, $\mathcal{P}(s_{t+1}\mid s_t, a_t)$ denotes the state-transition function, $\mathcal{R}(s_t, a_t)$ is the reward function, and $\gamma \in [0,1)$ is the discount factor. 
We model dynamic EMB--KV memory allocation as a Markov Decision Process (MDP) $(\mathcal{S}, \mathcal{A}, \mathcal{P}, \mathcal{R}, \gamma)$. At each decision epoch $t$, \allocator{} observes a runtime state $s_t \in \mathcal{S}$, takes an action $a_t \in \mathcal{A}$ to adjust the current allocation ratio $\alpha_t$, and the system transitions to a new state $s_{t+1} \sim \mathcal{P}(\cdot \mid s_t, a_t)$. The reward $r_t = \mathcal{R}(s_t, a_t)$ is computed from the serving performance observed over the subsequent decision window. 
The objective is to learn a policy $\pi(a_t \mid s_t)$ that maximizes the expected discounted return
\begin{equation}
\small
J(\pi) = \mathbb{E}_{\pi}\!\left[\sum_{t=0}^{\infty} \gamma^t r_t \right].
\end{equation}
This formulation captures the delayed effects of memory allocation decisions: adjusting $\alpha_t$ influences subsequent cache refill, eviction, and reuse, and its impact is reflected only after the system evolves to future states. We next describe the state, action, and reward design used to instantiate this MDP.

% At each decision epoch $t$, \allocator{} observes a runtime state $s_t \in \mathcal{S}$, chooses an action $a_t \in \mathcal{A}$, and the serving system transitions to a new state
% $s_{t+1} \sim \mathcal{P}(\cdot \mid s_t, a_t)$.
% The action $a_t$ specifies how the current EMB allocation ratio $\alpha_t$ should be adjusted, and the resulting reward $r_t = \mathcal{R}(s_t, a_t)$
% is computed from the serving performance observed over the subsequent decision window. The controller's objective is to learn a policy $\pi(a_t \mid s_t)$ that maximizes the expected discounted return
% {\small
% \begin{equation}
%\small
% J(\pi) = \mathbb{E}_{\pi}\!\left[\sum_{t=0}^{\infty} \gamma^t r_t \right]
% \end{equation}
% }
% This formula captures the delayed effect of memory allocation decisions: changing $\alpha_t$ influences later cache refill, eviction, and reuse, and these effects are reflected only after the serving system evolves to subsequent states. We next describe the state, action, and reward design used to instantiate this MDP.

\textbf{\emph{(1) State Space.}}
The state $s_t \in \mathcal{S}$ summarizes the recent workload, cache, and serving conditions that are most informative for the next memory-allocation decision. At each decision epoch $t$, \allocator{} observes a 7-dimensional state vector $s_t =(h_t, k_t, e_t, L_t, v_t, \alpha_t, \ell_t) \in \mathbb{R}^7$, where $h_t$, $k_t$, $e_t$, $L_t$, $v_t$, $\alpha_t$, and $\ell_t$ denote the hot-user ratio, KV hit rate, EMB hit rate, mean sequence length, SLO-violation rate, current EMB--KV memory allocation ratio, and P99 end-to-end latency, respectively. All features are normalized to $[0,1]$ before entering the allocation algorithm.

\begin{table}[t]
\footnotesize
\centering
\caption{State vector in \sysname{}.}
\label{tab:state}
\begin{tabular}{lp{5.5cm}}
\toprule
\textbf{Symbol} & \textbf{Definition} \\
\midrule
$h_t$        & Fraction of requests from hot users (\%) \\
$k_t$        & KV cache hit rate of recent window (\%) \\
$e_t$        & EMB cache hit rate of recent window (\%) \\
$L_t$        & Mean user interaction sequence length \\
$v_t$        & Fraction of latency SLO violations (\%) \\
$\alpha_t$   & Current EMB--KV memory allocation ratio (\%) \\
$p_t$        & Normalized P99 end-to-end request latency (\%) \\
\bottomrule
\end{tabular}
\end{table}

% \begin{figure}[t]
%   \centering
%   \begin{minipage}[t]{0.49\linewidth}
%     \centering
%     \includegraphics[width=\linewidth]{figures/fig_hot_user.png}
%     \caption{Request fraction across user popularity groups.}
%     \label{fig_hot_user}
%   \end{minipage}
%   \hfill
%   \begin{minipage}[t]{0.49\linewidth}
%     \centering
%     \includegraphics[width=\linewidth]{figures/fig_important_factors.pdf}
%     \caption{Feature importance for memory allocation decisions.}
%     \label{fig_important_factors}
%   \end{minipage}
% \end{figure}

\textbf{\emph{(2) Action Space.}}
The action $a_t \in \mathcal{A}$ is an incremental update to the current EMB allocation ratio, i.e., $a_t \equiv \Delta\alpha_t$. We use a small discrete action set
\begin{equation}
\small
\mathcal{A} = \{\pm0.06,\pm0.04,\pm0.02,0\}
\end{equation}
where positive values expand the EMB allocation, negative values shrink it, and zero leaves the current allocation unchanged. The next allocation ratio is then computed as
\begin{equation}
\small
\alpha_{t+1} = \mathrm{clip}(\alpha_t + \Delta\alpha_t,\; \alpha_{\min},\; \alpha_{\max})
\label{eq:alpha_update}
\end{equation}
where $\alpha_{\min}$ and $\alpha_{\max}$ bound the feasible EMB allocation range and prevent either cache from being fully starved. 
We use small incremental updates to avoid transient cache imbalance and latency spikes under live serving.

\textbf{\emph{(3) Reward Function.}}
After each action, the reward over the next decision window is defined as
\begin{equation}
\small
  r_t =
  \underbrace{\mathbb{E}_{\tau \sim \mathcal{T}_t}\!\bigl[\mathbf{1}[d_\tau \le \tau_{\mathrm{SLO}}]\bigr]}_{\text{QoS (SLO satisfaction)}}
  \;-\;
  \underbrace{\lambda \cdot \max\!\bigl(0,\, \ell_t - \tau_{\mathrm{SLO}}\bigr)}_{\text{P99 penalty (tail violation)}},
  \label{eq:reward}
\end{equation}
where the first term measures SLO satisfaction and the second penalizes P99 latency violation. 
Here, $\mathcal{T}_t$ is the set of requests in the current window and $d_\tau$ is the latency of request $\tau$. 
This formulation balances overall QoS and tail latency. Optimizing only SLO satisfaction may tolerate rare but severe latency spikes, whereas optimizing only P99 may overemphasize a small fraction of requests and neglect overall service quality. We set $\lambda = \frac{1}{\tau_{\mathrm{SLO}}}$ to keep both terms on a comparable scale.

\begin{figure}[t]
  \centering
  \includegraphics[width=0.8\linewidth]{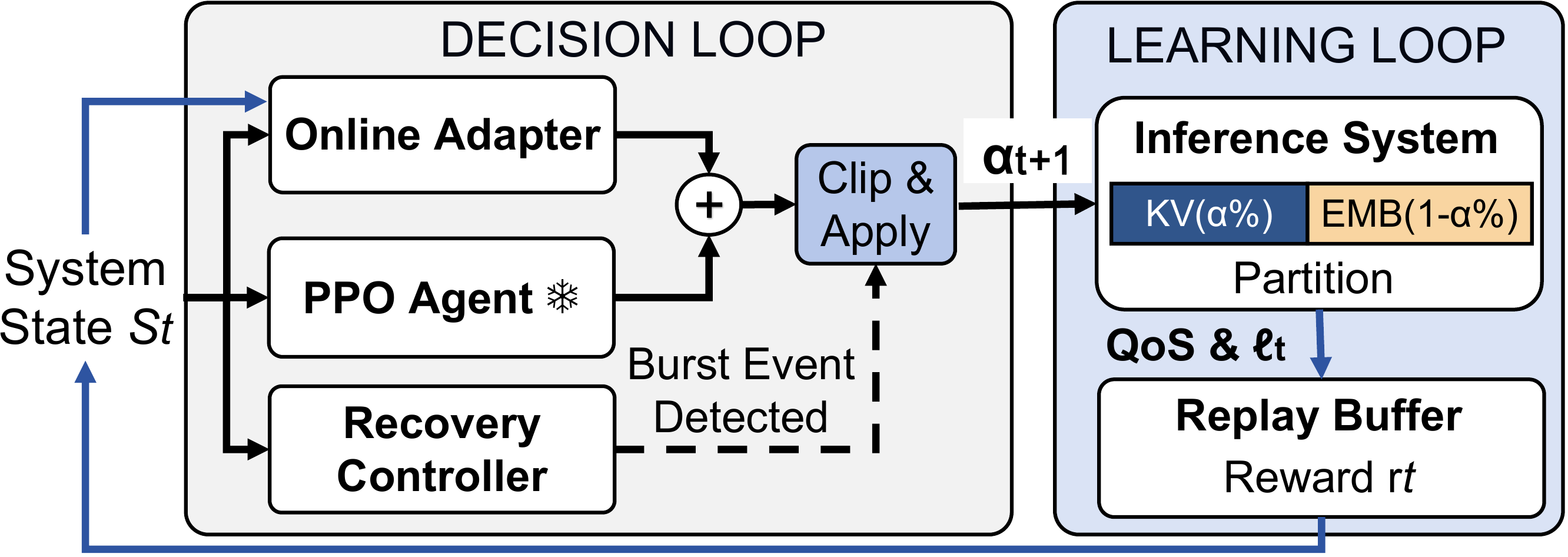}
  \caption{Two-loop controller architecture in \allocator{}}
  \label{fig_ai_model}
\end{figure}

\textbf{\allocator{} Architecture.}
To realize the above MDP in production serving, \sysname{} organizes the \allocator{} into two cooperating loops, as shown in~\autoref{fig_ai_model}. 
The \emph{decision loop} runs in the serving path: it consumes the current runtime state and produces the next EMB--KV allocation updates with low overhead. 
The \emph{learning loop} runs in the background maintenance path: it updates only the lightweight online adapter using recent serving experience while keeping the base policy fixed after offline training. 
% This design keeps online control efficient and stable while enabling \allocator{} to adapt to workload drift and burst events.

At each decision epoch (5s), the decision loop combines three runtime components: a base action from the PPO policy, a residual result from the Online Adapter, and a Recovery Controller override result when triggered. Concretely, the next EMB allocation ratio is computed as
\begin{equation}
\small
\alpha_{t+1} =
\mathrm{clip}\!\left(
\alpha_t + \Delta\alpha_{\mathrm{ppo}} + \Delta\alpha_{\mathrm{adapt}},
\; \alpha_{\min},\; \alpha_{\max}
\right),
\label{eq:decision_update}
\end{equation}
or applies the Recovery Controller's override during a burst event. Here, $\alpha_{\min}=0.1$ and $\alpha_{\max}=0.9$ ensure that neither cache is completely starved.

 \textbf{\emph{(1) PPO Agent (Base Frozen Policy).}}
    The PPO agent provides the dominant control signal $\Delta\alpha_{\mathrm{ppo}}$, mapping runtime state to effective memory allocation. PPO~\cite{schulman2017proximal} is used due to delayed effects: changes in $\alpha_t$ propagate through cache refill, eviction, and reuse before affecting end-to-end latency. 
    It is trained offline on historical traces over a discrete action space and converges within 100 episodes. We use a lightweight Actor--Critic with a two-layer MLP (64/128, $<25$K parameters). The policy is frozen after training to avoid instability and serving overhead.
    % The PPO agent provides the dominant control signal $\Delta\alpha_{\mathrm{ppo}}$, capturing the mapping from runtime state to an effective memory allocation under normal conditions. PPO~\cite{schulman2017proximal} is used here as the policy learner because memory-allocation decisions have delayed effects: a change in $\alpha_t$ propagates through later cache refill, eviction, and reuse before its impact is reflected in end-to-end latency. It is trained offline on historical traces over a discrete action space and converges within 100 episodes. We use a lightweight Actor--Critic model with a two-layer MLP (64/128, $<25$K parameters). The policy is frozen after training to avoid instability and serving-time overhead during deployment.

 \textbf{\emph{(2) Online Adapter (Residual Correction).}}
    This component compensates for offline–online mismatch by producing a residual correction $\Delta\alpha_{\mathrm{adapt}}$ over the base policy. It uses a lightweight residual model (3-layer MLP, $5{\to}16{\to}1$, zero-initialized). 
    The adapter is updated online from recent serving data. \allocator{} maintains an EMA of P99 latency, $\ell_{\mathrm{EMA}}$, and adjusts the learning rate as
    \begin{equation}
    \small
    \eta_t = \eta_0 \cdot \frac{|\ell_t - \ell_{\mathrm{EMA}}|}{\tau_{\mathrm{SLO}}}
    \end{equation}
    where $\eta_0 = 10^{-3}$. Each decision window's reward $r_t$ and latency $\ell_t$ are appended to a replay buffer (size 1K, FIFO), which bounds memory overhead and mitigates catastrophic forgetting under bursty workloads. 
    % Across five random seeds on a 4-hour trace with burst events, the online updates remain stable, with peak P99 deviation below 2.1\,ms and EMA-triggered rollbacks occurring fewer than twice per hour.

 \textbf{\emph{(3) Recovery Controller (Burst Handling).}}
    This component handles burst events, where stability is more important than fine-grained adaptation. It maintains an EMA baseline of P99 latency $\ell_{\mathrm{EMA}}$ and its standard deviation $\sigma_{\mathrm{EMA}}$ over the recent 10 decision epochs, and detects burst events when
    \begin{equation}
    \small
    \ell_t > \ell_{\mathrm{EMA}} + 3\sigma_{\mathrm{EMA}}
    \end{equation}
    Upon detection, the controller bypasses the normal update path and applies a conservative override, shifting memory toward the KV cache when the hot-user ratio exceeds $\rho^* = 0.2$. This fallback avoids contention on the H2D path during burst.

\begin{figure}[t]
\vspace{-2ex}
  \centering
  \begin{minipage}[t]{0.49\linewidth}
    \centering
    \includegraphics[width=0.9\linewidth]{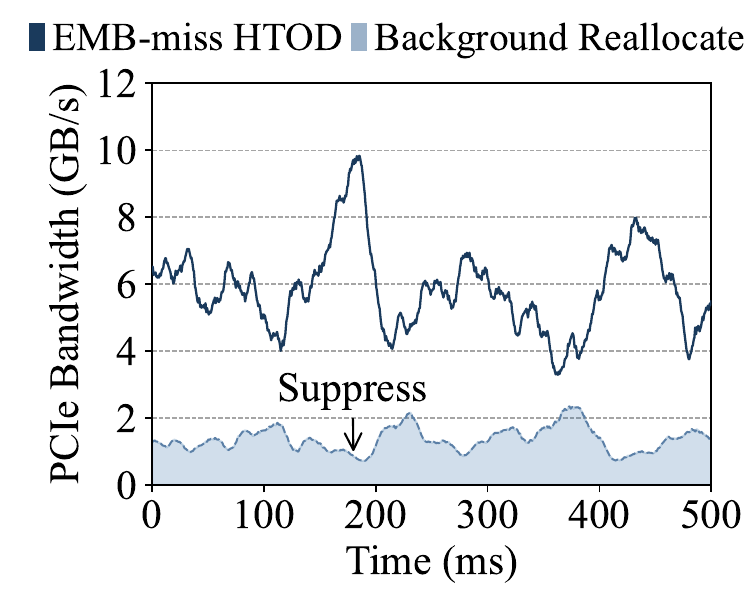}
    \caption{\sysname{} overlaps EMB HTOD and throttles background reallocation traffic}
    \label{fig_bandwidth}
  \end{minipage}
  \hfill
  \begin{minipage}[t]{0.49\linewidth}
    \centering
    \includegraphics[width=0.9\linewidth]{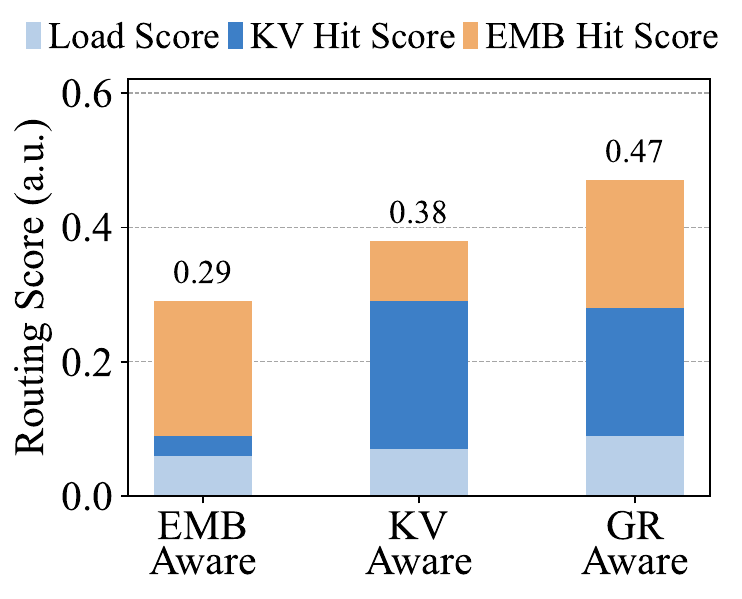}
    \caption{Routing score breakdown for three scheduling methods with different focuses}
    \label{fig_schedule_scores}
  \end{minipage}
\end{figure}

\subsection{Non-Disruptive Memory Allocation Adjustment}
\label{subsec_memory_management}
The Memory Manager faces a fundamental tension: \allocator{} operates on a 5s epoch and can issue allocation changes frequently, yet each change risks injecting H2D refill traffic that competes with latency-critical EMB-miss transfers. Resolving this tension requires decoupling logical boundary updates, which are metadata-only and incur microsecond-scale overhead, from physical refill, which is constrained by memory bandwidth.

Given a new EMB--KV memory allocation ratio from \allocator{}, the remaining challenge is to realize that update efficiently under live serving. 
In \sysname{}, changing the allocation incurs two distinct costs.
The first is a logical update to the EMB--KV memory boundary, which is lightweight and metadata-driven: \texttt{alloc\_page()} and \texttt{free\_page()} update page tables in less than 1\,$\mu$s and remain off the serving critical path. The second is a physical refill of the newly exposed capacity with useful cache content, which introduces background H2D traffic. In practice, a 4\% allocation update on an 80\,GB GPU corresponds to approximately 3.2\,GB of H2D refill traffic over PCIe, making refill rather than bookkeeping the dominant cost of online allocation adjustment.

Both EMB and KV caches occupy a shared HBM paged address space separated by a movable logical boundary. Above this virtual address layer, KV cache lookup is indexed by each user's exact interaction history: HSTU sequences are user-specific and share no common prefix across requests, precluding prefix-based or radix-tree indexing schemes. When free KV pages are exhausted or $\alpha$ increases, cold user blocks are reclaimed via metadata-only LRU eviction. Changing $\alpha$ updates this boundary; live KV blocks are not physically moved. When $\alpha$ increases, \sysname{} first reclaims cold, inactive KV blocks using LRU and returns their pages to the EMB pool. In-flight KV blocks are pinned and are never evicted. If there are not enough reclaimable KV pages, the EMB expansion is throttled/deferred to the next adjustment epoch. Thus, \texttt{alloc\_page()} succeeds only after safe KV reclamation, and eviction affects future KV hit rate but not correctness.

To apply allocation updates without degrading serving quality, \sysname{} decouples logical control from physical data movement and addresses two PCIe-bound challenges. \emph{(C1) Zero-copy allocation realization:} relocating live KV blocks would introduce prohibitive stalls on the inference path. \emph{(C2) Bandwidth interference:} H2D traffic induced by refill can interfere with latency-critical EMB-miss transfers on the critical path and amplify P99 latency.

\iffalse
\subsection{Non-Disruptive Memory Allocation Adjustment}
\label{subsec_memory_management}

Each node manages $\alpha$ at the node level; the 8 intra-node GPUs are abstracted as a unified HBM pool via NVLink (600,GB/s, $24\times$ PCIe). 
Memory reallocation is metadata-only: \texttt{alloc\_page()} and \texttt{free\_page()} update page tables in ${<}1\,\mu$s off the critical path. 
The dominant cost arises from data refill: a 4\% adjustment on an 80\,GB GPU triggers $\sim$3.2\,GB host-to-device (H2D) transfer over PCIe, which must be carefully controlled to avoid interfering with inference. 
Accordingly, \sysname{} decouples control and data paths and addresses 
two PCIe-bound challenges:
\emph{(C1) Zero-copy memory reallocation}: relocating live KV blocks would 
incur prohibitive stalls;
\emph{(C2) Bandwidth interference}: H2D traffic induced by reallocation can interfere with latency-critical EMB-miss transfers on the critical path.

\fi

\textbf{Embedding-only boundary adjustment (addresses C1).}
To avoid moving live KV state, \sysname{} enforces a strict invariant: only the EMB cache boundary is adjusted, while resident KV blocks remain physically immobile. The KV cache is organized as a paged block pool, similar to vLLM~\cite{kwon2023efficient}, where each block is managed through an explicit free list and indexed via page-table metadata. In contrast, the EMB cache is implemented as a contiguous LRU-managed slab backed by page-granularity allocation. When $\alpha$ increases, the EMB manager calls \texttt{alloc\_page()} to take free pages from the KV pool and reassign them to the EMB cache without touching resident KV entries. When $\alpha$ decreases, \texttt{evict\_lru\_entries()} reclaims EMB pages and returns them via \texttt{free\_page()} to the KV free list. Because all boundary changes are metadata-only and require no \texttt{cudaMemcpy}, allocation adjustment incurs \emph{zero pipeline stalls} regardless of scale.

\textbf{Background refill with bandwidth throttling (addresses C2).}
While boundary adjustment is lightweight, the newly allocated memory must still be refilled with useful cache content, which introduces background H2D traffic. Miss-path transfers are on the inference critical path and cannot be overlapped with attention execution (\S\ref{sec_motivation_challengs}), whereas proactive prefetch is a separate mechanism that speculatively warms the EMB cache during compute windows. To keep refill from interfering with serving-critical transfers, \sysname{} issues H2D movement via \texttt{cudaMemcpyAsync} on a dedicated high-priority CUDA stream (created with \texttt{cudaStreamCreateWithPriority}), decoupled from the default compute stream. \sysname{} further regulates refill bandwidth according to the current demand of EMB H2D traffic (Figure~\ref{fig_bandwidth}), throttling background allocation traffic when necessary to protect tail latency. Under controlled bandwidth usage, a typical 3.2\,GB H2D refill completes within approximately 1\,s, while the allocator operates at a slower 5\,s decision interval. This allows refill cost to be fully amortized without impacting end-to-end serving latency.

\subsection{KV-EMB-Aware Request Routing}
\label{subsec_scheduling}

\begin{algorithm}[t]
  \footnotesize
  \caption{KV-EMB Aware GR Request Routing}
  \label{algo:gr_schedule}
  \begin{algorithmic}[1]
    \State \textbf{Input:} request $R$, user $u$; nodes $\mathcal{N}$; miss costs $C_{\mathrm{kv}}, C_{\mathrm{emb}}, C_{\mathrm{ld}}$; overload threshold $\tau$; affinity bonus $\epsilon$
    \State \textbf{Output:} target node $n^*$
    \Statex \textbf{Router maintains per-node:} $\mathrm{Aff}[u]$: previous routed node per user; $\mathrm{KVM}[n]$: KV cache state HashMap $u\!\to\!\{0,1\}$; $\mathrm{EMBM}[n]$: EMB cache state HashMap; $\mathrm{Prof}[u]$:  top-$K$ hot shards for $u$; $l(n)$: load pressure
    \For{each request $R$ with user $u$}
        \State $n_{\mathrm{aff}} \leftarrow \mathrm{Aff}[u].\mathrm{node}$
             \hfill\Comment{Get Affinity Node}
      \For{each node $n \in \mathcal{N}$}
        \State $h_{\mathrm{kv}} \leftarrow \mathrm{KVM}[n].\mathrm{get}(u)$ \Comment{KV hit rate}
        \State $h_{\mathrm{emb}} \leftarrow |\mathrm{Prof}[u] \cap \mathrm{EMBM}[n]|\;/\;|\mathrm{Prof}[u]|$ \Comment{EMB hit rate}
        \State $\mathrm{score}(n) \leftarrow w_{\mathrm{kv}} h_{\mathrm{kv}} + w_{\mathrm{emb}} h_{\mathrm{emb}} + w_{\mathrm{ld}}(1-l(n)) + \epsilon$
      \EndFor
      \State $n^* \leftarrow \arg\max_{n}\,\mathrm{score}(n)$ \Comment{Best node}
      \If{$l(n^*) > \tau$}
        \State $n^* \leftarrow \arg\min_{n}\,l(n)$ \Comment{Overloaded: fall back to least-loaded}
      \EndIf
      \State Dispatch $R$ to $n^*$; Update $\mathrm{Aff}[u]\leftarrow n^*$; 
      Async update $\mathrm{KVM}[n^*]$ 
    \EndFor
  \end{algorithmic}
\end{algorithm}

Dynamic allocation introduces a routing dilemma: once nodes carry different $\alpha$ values, no single notion of locality, whether KV affinity or EMB coverage, is sufficient to score them. A KV-centric router overlooks the fact that nodes with small $\alpha$ have limited EMB cache capacity and are more likely to incur H2D transfer costs, while an EMB-centric router ignores KV residency and leads to recomputation. As a result, nodes are no longer interchangeable, and the same request may experience different latency depending on KV residency, embedding locality, and current load. To address this, \sysname{} adopts a \emph{user-aware} routing policy that evaluates each node based on the combined latency cost of both types of misses.

\textbf{Cost-Based Routing Score.}
Algorithm~\ref{algo:gr_schedule} summarizes the routing procedure. 
For each request $R$ from user $u$, the router scores every candidate node $n$ as
\begin{equation}
\small
\mathrm{score}(n)
=
w_{\mathrm{kv}} \cdot h_{\mathrm{kv}}
+
w_{\mathrm{emb}} \cdot h_{\mathrm{emb}}
+
w_{\mathrm{ld}} \cdot \bigl(1-l(n)\bigr)
+
b(n)
\label{eq:score}
\end{equation}
where $h_{\mathrm{kv}}\in\{0,1\}$ indicates whether node $n$ already holds reusable KV state for user $u$, $h_{\mathrm{emb}}\in[0,1]$ is the fraction of user $u$'s top-$K$ hot embedding shards resident on node $n$, and $l(n)\in[0,1]$ is the normalized node load. The affinity term is
\[
b(n)=\epsilon \cdot \mathbf{1}[n=n_{\mathrm{aff}}]
\]
which softly favors the user's last-routed node. The three main terms capture expected latency savings: a KV hit avoids sequence recomputation, higher EMB coverage reduces H2D fetch cost, and lower load reduces queueing delay.

\textbf{Weight and Threshold Configuration.}
The routing weights are derived from offline-profiled miss costs rather than tuned manually. Let $C_{\mathrm{kv}}$, $C_{\mathrm{emb}}$, and $C_{\mathrm{ld}}$ denote the physical costs of a KV miss, an embedding miss, and marginal queueing delay, respectively, and let $Z = C_{\mathrm{kv}} + C_{\mathrm{emb}} + C_{\mathrm{ld}}$.
Then the routing weights are normalized as $w_i = C_i / Z$, so that
$w_{\mathrm{kv}} + w_{\mathrm{emb}} + w_{\mathrm{ld}} = 1$.
In our implementation, the overload threshold is $\tau = 0.85$, the affinity bonus is $\epsilon = 0.05$, and the shard-profile size is $K=20$, which covers at least 95\% of per-user embedding accesses while remaining insensitive for $K\in[10,50]$. If the highest-scoring node exceeds the overload threshold, the router falls back to the least-loaded node.

\textbf{Residency Tracking.}
To support routing efficiently, \sysname{} maintains lightweight per-node metadata for both KV and EMB locality. KV residency is tracked at user granularity as a hash map from user ID to a presence bit, while EMB residency is tracked at shard granularity as a set of resident embedding-shard identifiers. For each request, $h_{\mathrm{emb}}$ is computed by intersecting the node's EMB-residency set with the user's precomputed top-$K$ hot shards, which is an $O(K)$ operation. These structures are updated asynchronously off the routing critical path and remain inexpensive in practice: the KV-residency table occupies under 24\,KB per node, and at $\alpha=0.5$ each node typically holds approximately $3\text{K}$--$6\text{K}$ KV entries.

\textbf{System-Level Effect.}
By scoring nodes using both cache perspectives and load, \sysname{} avoids EMB- or KV-only routing myopia and exploits heterogeneous memory conditions from dynamic EMB--KV allocation. As shown in \autoref{fig_schedule_scores}, our scheduler achieves a higher aggregate routing score (0.47) than EMB-only (0.29) or KV-only (0.38) by jointly optimizing all cost components. This closes the loop: \allocator{} adapts per-node EMB--KV allocation, the memory manager enforces SLOs, and the router dispatches requests accordingly.

Together, the controller, memory manager, and router form a closed runtime loop for dynamic EMB--KV management in \sysname{}. \allocator{} chooses the memory allocation, the memory manager applies it under live traffic, and the router exploits the resulting per-node cache conditions for request placement. Section~\ref{sec_evaluation} evaluates how this coordinated design improves end-to-end latency, SLO satisfaction, and scalability.
\section{EVALUATION}
\label{sec_evaluation}

\subsection{Experimental Setup}
\label{sec_eval_config}
\textbf{System Setup.}
\sysname{} is implemented on top of the open-source HSTU framework~\cite{zhai2024actions}, a production-grade generative recommender architecture with wide industrial adoption. Requests are served via continuous batching at user-request granularity~\cite{yu2022orca}. We use FBGEMM~\cite{fbgemm} for optimized embedding kernels and NCCL~\cite{nccl} for collective communication. 
All experiments are conducted on a 32-node cluster. Each node is equipped with 8$\times$ NVIDIA A100 GPUs (80\,GB HBM2e), a dual-socket AMD EPYC 9684X CPU with a total of 2\,TB DDR5 memory, and a 200\,Gbps InfiniBand HDR interconnect. This setup ensures that the dominant bottlenecks are on-node memory allocation and routing rather than network bandwidth.

\textbf{Model Configurations.}
We evaluate two HSTU configurations to cover both lightweight and heavyweight serving regimes: a 3-layer model corresponding to the ranking stage and a 6-layer model corresponding to the retrieval stage. Both models use 512-dimensional item embeddings. This setup allows us to evaluate \sysname{} under different compute and memory intensities while keeping the serving stack unchanged.

%Requests are served via continuous batching at user-request granularity~\cite{yu2022orca}. 

\textbf{Baselines.}
We compare \sysname{} against two groups of baselines under a 30\,ms P99 SLO. The first group evaluates end-to-end serving systems that optimize EMB and KV in isolation rather than jointly. 
\begin{itemize}[leftmargin=*]
    \item \emph{KV-Opt} represents KV-centric serving systems~\cite{qin2025mooncake,sun2026bat,sun2025xgr} that optimize KV reuse and scheduling with $\alpha=0$, treating embedding access as secondary.
    \item  \emph{KV-EMB-Opt} extends the KV-centric serving stack with ML-guided EMB prefetching~\cite{ren2025machine}, but still uses a \emph{fixed} EMB--KV memory allocation during serving. To give it the strongest possible static configuration, we select its allocation ratio $\alpha^*$ offline by exhaustive grid search over $\{0.0, 0.1, \dots, 1.0\}$ and use the value that yields the lowest P99 latency on the target workload.
\end{itemize}

The second baseline group isolates the dynamic memory allocation controller, while keeping the serving infrastructure fixed. We compare against three classes of methods. 
\begin{itemize}
    \item \emph{Classical controllers:} PID~\cite{astrom1995pid}, Bayesian Optimization (BO)~\cite{shahriari2015taking}, and MILP~\cite{gurobi2021}. 
    \item \emph{Predictive learning-based methods:} XGBoost~\cite{chen2016xgboost} and LSTM~\cite{yu2019review}, which estimate the best allocation from observed runtime state. 
    \item \emph{PPO-only:} our base PPO controller with the Online Adapter and Recovery Controller disabled, isolating the contribution of online correction and burst handling.
\end{itemize}

\begin{table}[t]
\centering
\footnotesize
\caption{Datasets used for evaluation}
\label{tab_datasets_all}
\begin{tabular}{lcc}
\hline
\textbf{Dataset} & \textbf{Users} & \textbf{Top 5\% User Req. Ratio} \\
\hline
Taobao             & 0.98M & 18\% \\
Amazon-Video-Games & 2.76M    & 24\% \\
Amazon-Books        & 10.3M   & 38\%   \\
\hline
\end{tabular}
\end{table}

\textbf{Datasets and Workload Generation.}
We use three real-world datasets (\autoref{tab_datasets_all}): 
Taobao~\cite{taobao} (e-commerce clicks), Amazon Video-Games and 
Amazon Books~\cite{hou2024bridging} (product reviews), covering 
low- and high-cardinality embedding regimes. 
To approximate production-scale serving pressure, we replay request traces based on production workloads from Company Alibaba 
%\footnote{Company X is anonymized due to the double-blind review policy.}
, request arrival rates are drawn from each user's historical access frequency. 
To stress the system, we scale user profiles to 5K–15K tokens and retrieve 100 candidate items per request, following recent long-sequence GR settings~\cite{chai2025longer,guan2025make,zhou2026gems}. 
To reflect 
production-scale embedding memory pressure~\cite{yu2026near, lian2022persia,zhao2023recd}, we enlarge 
embedding tables to 5\,TB using NVIDIA DLRM scripts~\cite{dlrm_datasets}, 
while preserving each dataset's original Zipfian ID-frequency distribution. This is important because the Zipf skew directly affects HBM miss rates and the sensitivity of EMB--KV memory allocation.

\textbf{Workload Regimes.}
We evaluate three workload regimes that stress different aspects of the system. \emph{Steady} uses uniform arrivals with a stable hot-user ratio. \emph{Trend} gradually increases the hot-user ratio over time, testing whether \allocator{} can track a drifting optimum. \emph{Burst} injects sudden hot-user spikes through Poisson arrivals lasting 3--5 decision windows, stressing both tail-latency robustness and the burst-handling logic in \sysname{}.

\textbf{Metrics.}
We report two classes of metrics. \emph{(1) End-to-end serving performance:} P99 latency (ms) and QoS rate, defined as the fraction of requests meeting the 30\,ms SLO. \emph{(2) Allocation quality:} the oracle ratio gap defined as 
$\lvert \alpha_t - \alpha^*_t \rvert$, where $\alpha^*_t$ is the 
epoch-level offline optimum obtained by exhaustively evaluating 
$\alpha \in \{0.0, 0.01, \dots, 1.0\}$ using the full workload trace of epoch $t$ and selecting the value minimizing mean P99 latency. The gap measures how closely the policy tracks it. For Burst events, $\alpha^*_t$ is computed independently per epoch to capture rapidly shifting optima.

\begin{figure}[t]
  \centering
  \includegraphics[width=0.8\linewidth]{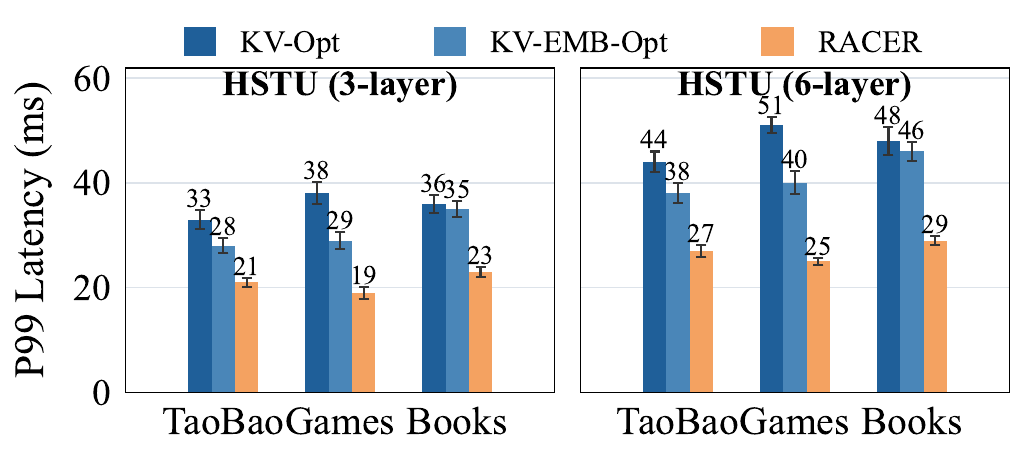}
  \caption{Overall latency P99 of three methods across three production-scale datasets on two model depths}
  \label{fig_overall_latency}
\end{figure}

\subsection{Overall End-to-End Performance}
\label{sec_eval_performance}

\textbf{P99 Latency.}
~\autoref{fig_overall_latency} reports end-to-end P99 latency across datasets and HSTU model configurations. \sysname{} consistently satisfies the 30\,ms P99 SLO across all datasets, model depths, and workload regimes (95\% confidence intervals from five independent runs), whereas neither baseline does so reliably. {KV-Opt} already exceeds the SLO under steady load, reaching 33--38\,ms for the 3-layer model and 44--51\,ms for the 6-layer model, indicating that KV-centric optimization alone is fundamentally insufficient once embedding access becomes a first-order bottleneck. 
{KV-EMB-Opt} improves over {KV-Opt} by combining KV optimization with EMB prefetching and a fixed offline-tuned allocation ratio, but still cannot meet the SLO consistently, yielding 28--35\,ms for the 3-layer model and 38--46\,ms for the 6-layer model.

In contrast, \sysname{} remains within the SLO across all settings, achieving 19--23\,ms for the 3-layer model and 25--29\,ms for the 6-layer model. Compared with the strongest static baseline, {KV-EMB-Opt}, \sysname{} reduces P99 latency by 25--34\% on the 3-layer model and 28--38\% on the 6-layer model. 
These gains confirm that jointly adapting EMB--KV memory allocation at runtime is more effective than any fixed configuration, even when that configuration is chosen offline to be as favorable as possible. 
The gains are particularly pronounced on Amazon Books, where the hot-user ratio reaches 38\%, because higher hot-user concentration intensifies EMB--KV contention and makes static allocation increasingly ineffective. Finally, the small run-to-run variance (within 1.2\,ms) indicates that \sysname{} stabilizes tail latency rather than merely shifting the average case.

\begin{figure}[t]
  \centering
  \includegraphics[width=0.9\linewidth]{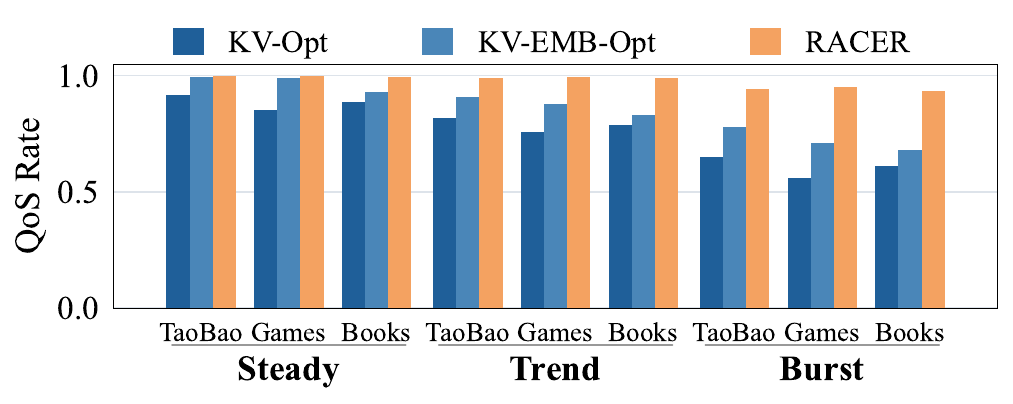}
  \caption{SLO satisfaction rate of three methods across three production-scale datasets under 30\,ms latency SLO.}
  \label{fig_overall_qos}
\end{figure}

\textbf{SLO Satisfaction Rate.}
~\autoref{fig_overall_qos} reports the fraction of requests that meet the 30\,ms SLO under the three workload regimes. Under \emph{Steady} traffic, the gap between \sysname{} and \emph{KV-EMB-Opt} is relatively modest: \sysname{} achieves at least 99.6\% SLO satisfaction, while \emph{KV-EMB-Opt} reaches 92.8--99.4\%. This is expected, since a well-tuned fixed allocation can remain close to optimal when the workload remains stable. Under \emph{Trend}, however, the gap widens: \emph{KV-EMB-Opt} drops to 83--91\% as the hot-user ratio drifts away from its design point, whereas \sysname{}
maintains at least 98.9\% by tracking the changing EMB--KV optimum online.

The largest difference appears under \emph{Burst} traffic. Static methods degrade sharply to 56--78\% SLO satisfaction because sudden hot-user spikes quickly invalidate their fixed allocation, whereas \sysname{} sustains 93.5--95.3\% through the combined effect of adaptive allocation, non-disruptive adjustment, and burst-aware recovery control. Across datasets, \sysname{}'s QoS variation under \emph{Burst} remains within 1.8\%, compared with 12--22\% for static baselines. This result shows that \sysname{} not only lowers tail latency, but also makes system behavior significantly more robust to workload volatility and reduces the need for dataset-specific tuning.

\subsection{Ablation Study}
\label{sec_eval_ablation}

\begin{figure}[t]
    \centering
    \includegraphics[width=0.95\linewidth]{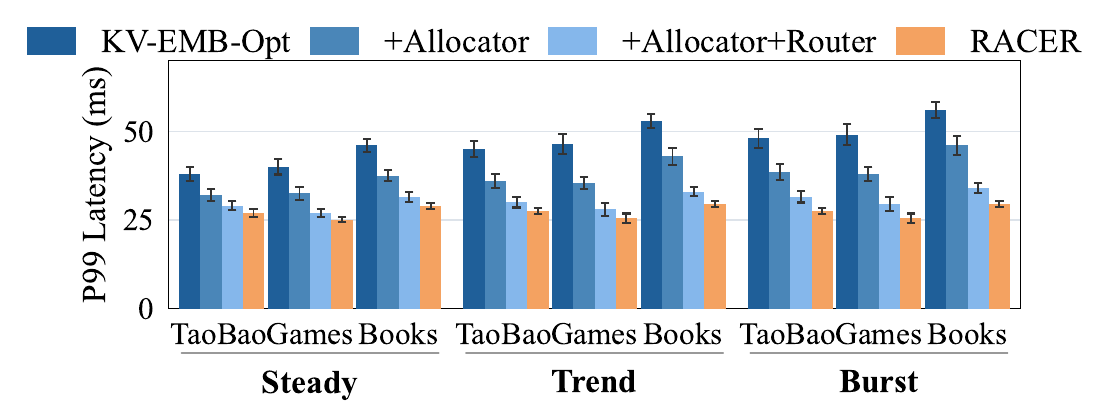}
    \caption{System-level ablation of \sysname{} }
    \label{fig_ablation_latency}
\end{figure}

\begin{figure}[t]
\centering
    \includegraphics[width=0.7\linewidth]{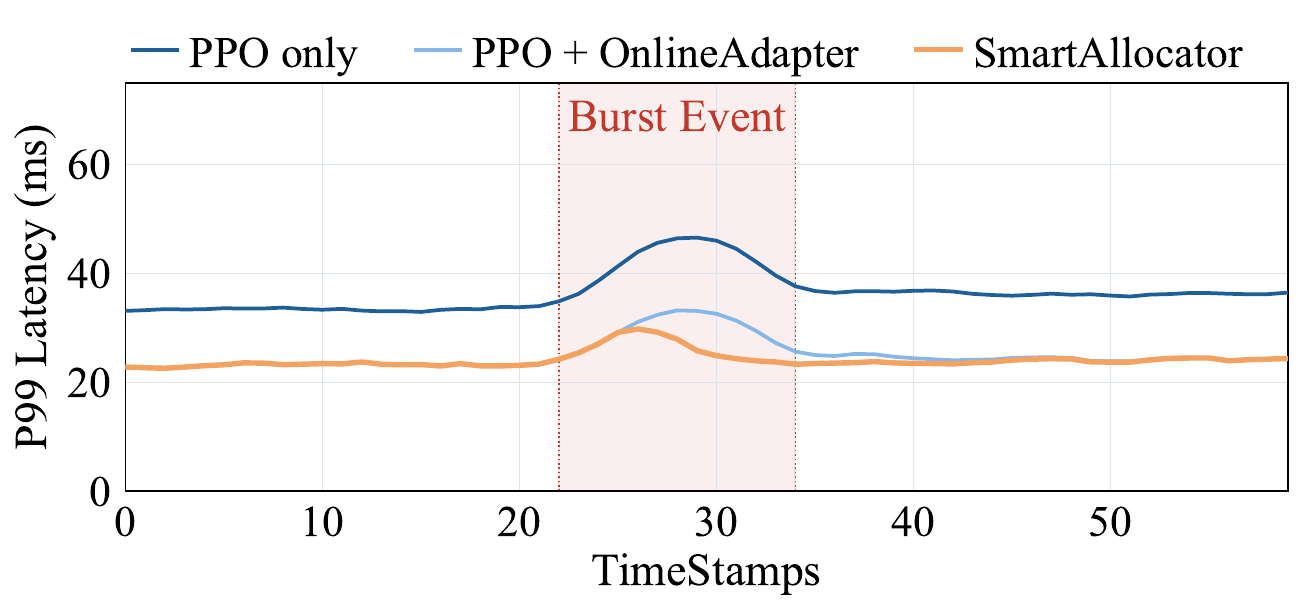}
    \caption{Allocator-internal ablation study}
    \label{fig_ablation_Allocator}
\end{figure}

\textbf{System-Level Ablation.}
\autoref{fig_ablation_latency} quantifies the end-to-end contribution of \sysname{}'s two system-level components: \allocator{} and the EMB--KV-aware request router. 
Adding the allocator alone reduces P99 latency by $\sim$20\% compared to baseline with fixed $\alpha^*$, confirming that adaptive EMB--KV memory allocation already resolves a substantial portion of the dual-bottleneck problem.
This gain is largest under Burst workloads, where fixed allocation under-provisions the active cache. Adding the router yields a further 10--25\% reduction by concentrating KV residency for repeated users while preserving embedding locality, resulting in a more cache-friendly request distribution across nodes.

\textbf{Allocator-Internal Ablation.}
\autoref{fig_ablation_Allocator} ablates the allocator components under a representative burst. PPO-only maintains $\sim$33 ms steady-state latency but reacts slowly, with P99 peaking at $\sim$46 ms before gradual recovery. Adding OnlineAdapter lowers steady-state latency to $\sim$23 ms and peak to $\sim$34 ms, but burst remains sluggish; the full \allocator{} maintains the same baseline while the Recovery Controller caps the peak at $\sim$31 ms and quickly collapses the tail. 

% Adding OnlineAdapter reduces steady-state to $\sim$23 ms and the peak to $\sim$34 ms, yet recovery remains sluggish. The full \allocator{} shares the same steady-state and rising phase as OnlineAdapter, and the Recovery Controller then intervenes mid-burst, capping the peak at $\sim$31 ms and collapsing the tail within a few epochs.

\textbf{Control Overhead.}
These gains are achieved with low runtime overhead. OnlineAdapter training increases peak CPU utilization from 20\% to 46\%, with an average utilization of 32\%, and runs entirely in the background without affecting request serving. On the router side, the affinity table and KV residency table together occupy under 100\,KB per node at 32-node scale, indicating that the metadata required for EMB--KV-aware routing is negligible.

\subsection{Sensitivity Study}
\label{sec_eval_sensitivity}

\begin{table}[t]
\footnotesize
\centering
\renewcommand{\arraystretch}{0.7}
\caption{Comparison of $\alpha$ control strategies.
  \textbf{Dec.~Time}: time to produce one allocation decision;
  \textbf{Opt.~Gap}: mean deviation from offline-optimal $\alpha^*$
  across three workload scenarios.}
\label{tab:ablation_strategy}
\begin{tabular}{lcccc}
\toprule
\textbf{Method} & \textbf{Dec.~Time} 
  & \multicolumn{3}{c}{\textbf{Opt.~Gap}} \\
\cmidrule(lr){3-5}
 & & \textbf{Steady} & \textbf{Trend} & \textbf{Burst} \\
\midrule
Static-$\alpha$=0.1 & \multirow{5}{*}{0\,s} 
                    & 0.292 & 0.341 & 0.331 \\
Static-$\alpha$=0.3 && 0.092 & 0.141 & 0.131 \\
Static-$\alpha$=0.5 && 0.108 & 0.061 & 0.072 \\
Static-$\alpha$=0.7 && 0.308 & 0.259 & 0.269 \\
Static-$\alpha$=0.9 && 0.508 & 0.459 & 0.469 \\
\midrule
PID    & $\sim$1.2\,s & 0.094 & 0.108 & 0.167 \\
XGBoost& $\sim$3\,ms  & 0.053 & 0.091 & 0.201 \\
BO     & $\sim$7\,s   & 0.061 & 0.089 & 0.174 \\
MILP   & $\sim$36\,s  & 0.044 & 0.068 & 0.161 \\
\midrule
LSTM             & $\sim$5\,ms & 0.051 & 0.083 & 0.156 \\
LSTM+OFT         & $\sim$6\,ms & 0.043 & 0.067 & 0.128 \\
LSTM+OFT+RC      & $\sim$6\,ms & 0.041 & 0.069 & 0.051 \\
\midrule
PPO Only         & 28\,$\mu$s  & 0.043 & 0.051 & 0.042 \\
\textbf{\allocator{}} & \textbf{32\,$\mu$s} 
  & \textbf{0.019} & \textbf{0.023} & \textbf{0.031} \\
\bottomrule
\end{tabular}
\end{table}

\textbf{Comparison of $\boldsymbol{\alpha}$-control strategies.}
Table~\ref{tab:ablation_strategy} compares \allocator{} against alternative controllers using two metrics: the gap to the oracle offline-optimal allocation ratio $\alpha^*$ and the time required to produce one allocation decision.

No static $\alpha$ performs competitively across regimes, confirming that fixed allocation is fundamentally inadequate under changing workloads.
Classical non-RL methods suffer from \emph{objective mismatch}: PID, BO, MILP, and XGBoost optimize single-step objectives, yet allocation changes influence P99 latency only after cache warm-up and eviction effects propagate across multiple epochs. As a result, these methods either react too slowly or overreact under bursty workload shifts, yielding Burst gaps between 0.161 and 0.201. Their decision latency is also substantial in several cases, e.g., 1.2\,s for PID, 7\,s for BO, and 36\,s for MILP, making them impractical for online control.
LSTM-based methods partially mitigate this mismatch through temporal modeling, and LSTM+OFT+RC further improves Burst behavior by adding online fine-tuning and burst handling, reducing the Burst gap to 0.051. However, its aggressive overrides inflate the Steady and Trend gaps to 0.041--0.069, showing that temporal prediction alone does not fully resolve the trade-off between responsiveness and stable long-horizon control. 
RL-based methods, including PPO-only and \allocator{}, improve both allocation quality and responsiveness. PPO-only reduces the gap to 0.043, 0.051, and 0.042 under Steady, Trend, and Burst, respectively, at only 28\,$\mu$s per decision. \allocator{} further reduces this to 0.019, 0.023, and 0.031 while increasing decision latency slightly to 32\,$\mu$s.

\begin{figure}[t]
  \centering
  \includegraphics[width=0.9\linewidth]{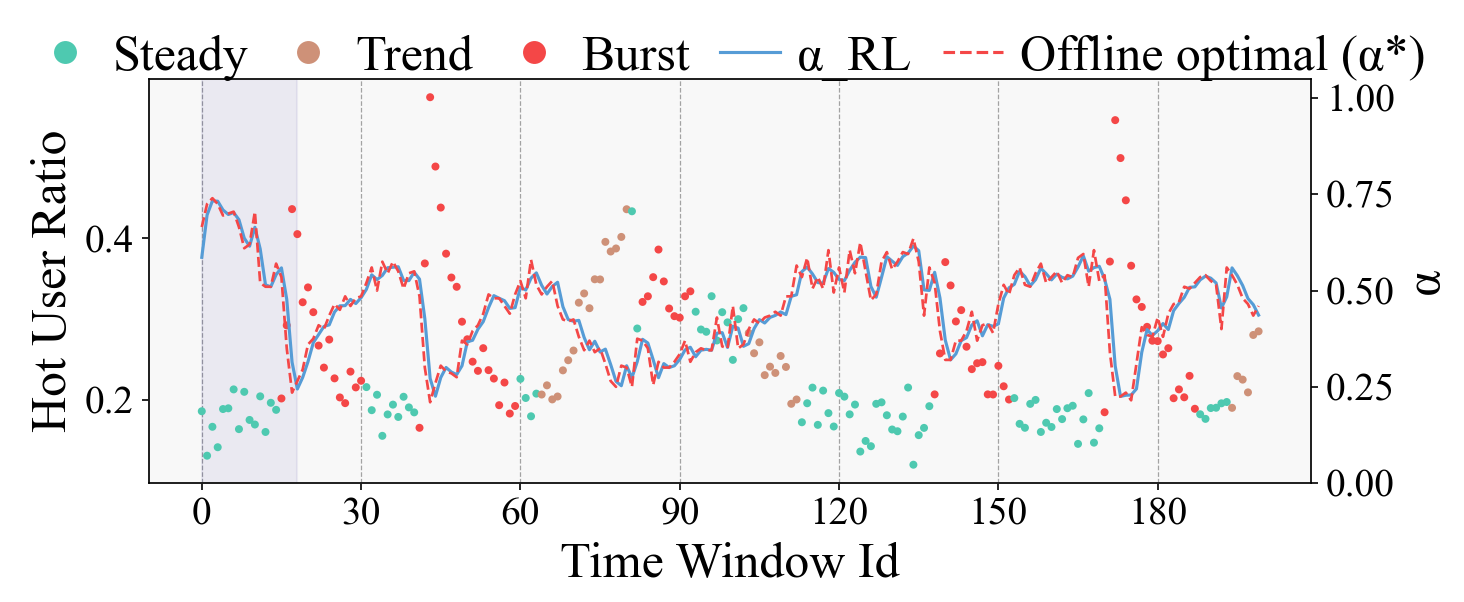}
  \caption{RL-controlled $\alpha_{RL}$ versus offline optimal $\alpha^*$ over a four-hour window (Amazon-Books). Scatter points denote hot-user request regimes:
  \textcolor[RGB]{82,188,163}{\textbf{Steady}},
  \textcolor[RGB]{188,143,90}{\textbf{Trend}}, and
  \textcolor[RGB]{220,50,50}{\textbf{Burst}}.}
  \label{fig_partition_ratio_change}
\end{figure}

\textbf{Tracking offline-optimal $\boldsymbol{\alpha^*}$.}
Figure~\ref{fig_partition_ratio_change} shows how the RL-controlled allocation ratio $\alpha_{\mathrm{RL}}$ evolves over time relative to the offline optimum. After the warm-up region (shaded), $\alpha_{\mathrm{RL}}$ closely tracks $\alpha^*$ with mean absolute error 0.023, despite having no future knowledge of the incoming workload. The learned policy responds differently across operating regimes: under Steady load ($t\!\approx\!90$, hot-user ratio $\approx 0.20$), it raises $\alpha_{\mathrm{RL}}$ to 0.51 to favor EMB locality (EMB/KV hit rates: 55\% / 21\%); under Trend ($t\!\approx\!130$, hot-user ratio $\approx 0.32$), it lowers the ratio to 0.38 as KV pressure grows (50\% / 18\%); under Burst ($t\!\approx\!175$, hot-user ratio $\approx 0.50$), it further drops to 0.26, trading a 9\% EMB-hit reduction for a 25\% KV-hit gain to avoid SLO violations.
% This behavior reflects asymmetric degradation: KV performance drops sharply once capacity falls below its working set, while EMB degrades more gradually; as the hot-user ratio increases, the KV working set expands super-linearly, amplifying the cost of each KV miss on the critical path.
This reflects asymmetric degradation: KV performance drops sharply below its working-set capacity while EMB degrades gradually; as the hot-user ratio grows, the KV working set expands super-linearly, amplifying each KV miss on the critical path.

\begin{figure}[t]
  \centering
  \includegraphics[width=0.88\linewidth]{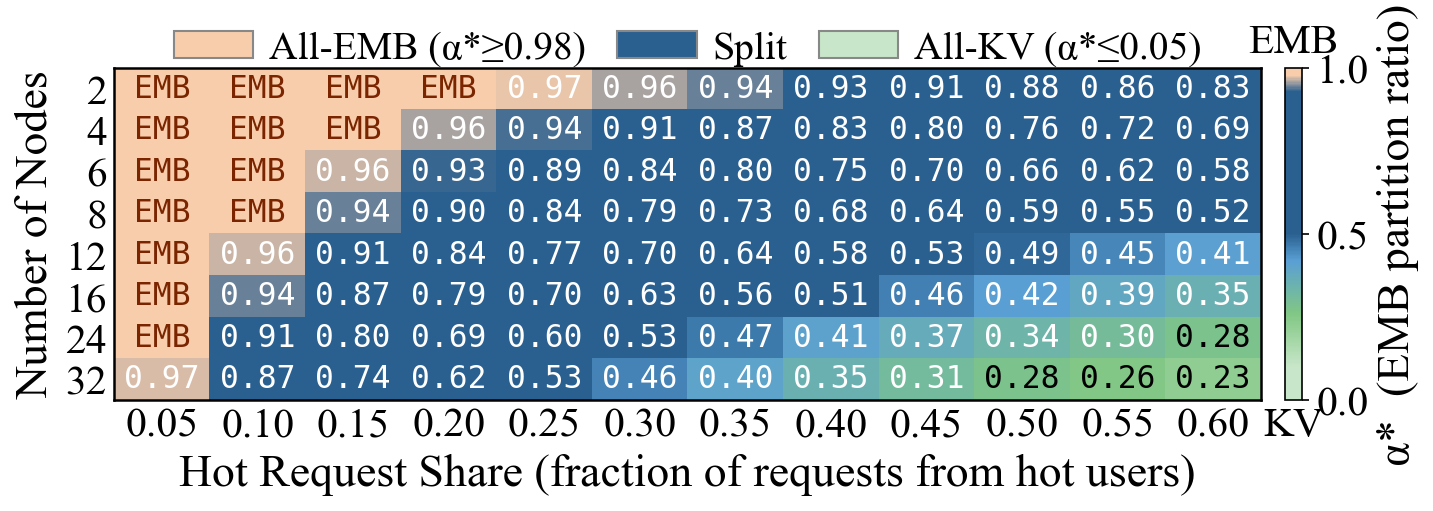}
  \caption{Average $\alpha_{\text{RL}}$ under varying cluster scale and hot-request share. Color denotes regime: \textcolor{orange!70!black}{All-EMB}, \textcolor{blue!70!black}{Split}, \textcolor{green!60!black}{All-KV}.}
  \label{fig_regime_map}
\end{figure}

\begin{figure}[t]
  \centering
  \includegraphics[width=0.9\linewidth]{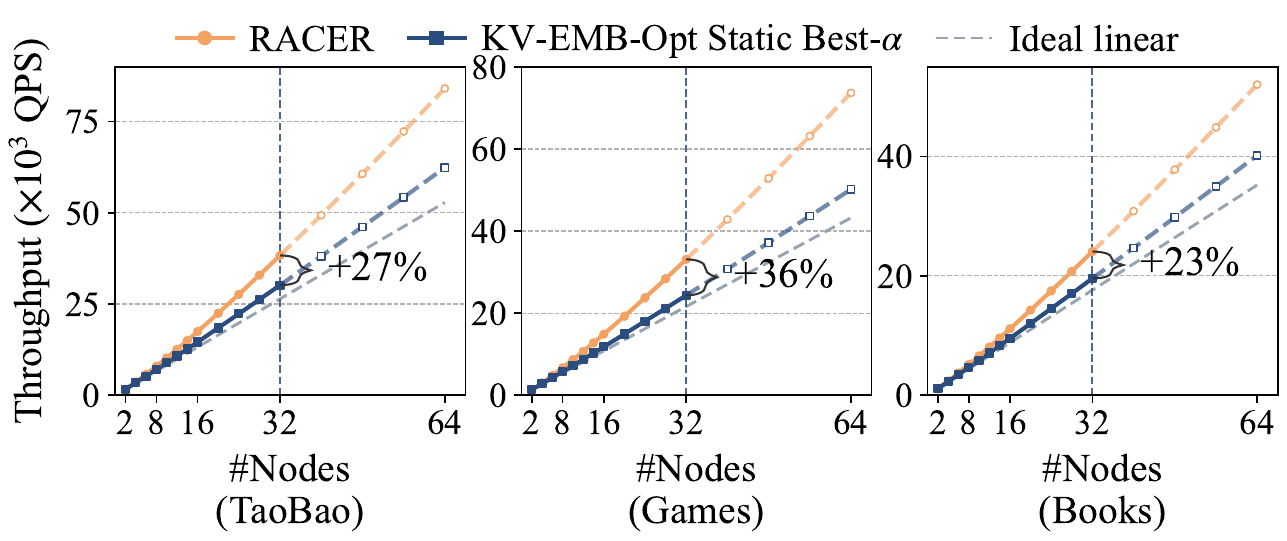}
  \caption{Weak-scaling throughput (QPS) on three datasets as node count scales from 2 to 32.}
  \label{fig_scalability}
\end{figure}

\textbf{Sensitivity to User Hotness and Cluster Scale.}
Figure~\ref{fig_regime_map} sweeps cluster size (2--32 nodes) and hot-request share (0.05--0.60) while fixing the underlying hot-user population ratio. The heatmap reports the average $\alpha_{\mathrm{RL}}$ selected by \allocator{}. First, as hot-request share increases, the controller consistently reduces EMB allocation to preserve more KV capacity. Second, increasing cluster size further lowers $\alpha_{\mathrm{RL}}$ because embedding tables are sharded across more nodes, reducing each node's EMB working set and shifting the balance toward KV. Overall, the system transitions from an EMB-dominated regime (2 nodes, hot-request share 0.05, $\alpha_{\mathrm{RL}}=1$) to a KV-dominated regime (32 nodes, hot-request share 0.60, $\alpha_{\mathrm{RL}}\approx 0.23$), showing that \allocator{} adapts across regimes where no static allocation is appropriate.

\subsection{Scalability}
% We evaluate weak scaling by fixing the embedding data per node and proportionally increasing the total dataset pressure as the cluster grows from 2 to 32 nodes, with all nodes sharing identical hardware configurations (Weak scaling). Under this setup, ideal linear scaling implies constant per-node throughput regardless of cluster size. As shown in ~\autoref{fig_scalability}, \sysname{} maintains near-linear throughput growth across all three datasets, closely tracking the ideal baseline. At 32 nodes, \sysname{} outperforms KV-EMB-Opt by 27\%, 36\%, and 23\% on Taobao, Games, and Books, respectively. The performance gap widens with cluster size, indicating that KV-EMB-Opt incurs growing cross-node coordination overhead as the total data volume scales, while \sysname{}'s workload-aware allocation keeps per-node embedding traffic balanced and minimizes unnecessary inter-node communication.

\textbf{Weak-Scaling Throughput.}
% We evaluate weak scaling by fixing the embedding data per node and proportionally increasing total dataset pressure as the cluster grows from 2 to 32 nodes, with all nodes using identical hardware. Under this setup, ideal linear scaling corresponds to constant per-node throughput as the system size increases. Figure~\ref{fig_scalability} shows that \sysname{} maintains near-linear throughput growth across all three datasets, closely tracking the ideal baseline. At 32 nodes, \sysname{} outperforms KV-EMB-Opt by 27\%, 36\%, and 23\% on Taobao, Games, and Books, respectively. The gap widens with cluster size, indicating that static KV/EMB optimization incurs increasing cross-node coordination overhead as data volume grows, whereas RACER's workload-aware memory allocation and routing keep per-node embedding traffic balanced and reduce unnecessary inter-node communication. 
We evaluate weak scaling by fixing the embedding data per node and proportionally increasing total dataset size as the cluster grows from 2 to 32 nodes. Ideal linear scaling corresponds to constant per-node throughput. Figure~\ref{fig_scalability} shows that \sysname{} maintains near-linear throughput growth across all three datasets. At 32 nodes, \sysname{} outperforms KV-EMB-Opt by 27\%, 36\%, and 23\% on three datasets respectively. The performance gap widens with cluster size, as static KV/EMB optimization incurs increasing cross-node coordination overhead, whereas \sysname{}'s workload-aware memory allocation and routing keep per-node embedding traffic balanced and reduce unnecessary inter-node communication.

\textbf{Hardware Generalizability.}
The dual bottleneck in GR serving is governed by the ratio $F_{\mathrm{gpu}}/B_{\mathrm{pcie}}$, which compares KV recomputation cost ($\propto L^2 / F_{\mathrm{gpu}}$) against EMB-miss transfer cost ($\propto 1 / B_{\mathrm{pcie}}$). This ratio remains stable across recent GPU generations: A100 yields 12.5, while H100 and H200 yield 15.5, a modest increase of at most 24\% that keeps both costs first-order bottlenecks. Although H200 provides larger HBM capacity (141\,GB/GPU), it does not remove the zero-sum EMB--KV trade-off: hot embeddings still exceed HBM by more than an order of magnitude, and KV tensors at sequence lengths 8K--15K (60--184\,MB each) still limit an 8-GPU node to roughly 3K--6K active users. These observations suggest that EMB--KV contention is structural rather than A100-specific, and that RACER's design should remain relevant across future GPU generations. This bottleneck observation is consistent with the recent MLperf's DLRMv3 benchmark (HSTU baseline)~\cite{mlcommons_dlrmv3_2026}, where H100-based evaluations still report substantial latency contributions from both operations (EMB=30\% vs. Dense=48\%) using only one embedding table, aligning with our profiling results in~\autoref{fig_hstu_breakdown}.

% \textbf{Hardware Generalizability Analysis.}
% The dual bottleneck is governed by $F_{\text{gpu}}/B_{\text{pcie}}$, comparing KV recomputation ($\propto L^2/F_{\text{gpu}}$) and EMB-miss transfer ($\propto 1/B_{\text{pcie}}$). This ratio remains stable across generations: A100 yields $12.5$, H100~\cite{nvidia_h100} and H200~\cite{nvidia_h200}$\,$ yield $15.5$, a modest ${\leq}24\%$ increase that keeps both as first-order costs. Although H200’s larger HBM (141\,GB/GPU) raises capacity, it does not remove the zero-sum trade-off: hot embeddings still exceed HBM by an order of magnitude and KV tensors at $L=8$K--15K (60--184\,MB) limit even an 8-GPU node to ${\sim}3k-6k$ users. Thus, EMB--KV contention remains structural across hardware generations.

\section{RELATED WORK}
% embedding优化方法  kv cache优化方法
\textbf{EMB and KV Optimizations for Recommendation Models.} 
While prior work has explored cache optimizations, improving either component in isolation may boost its individual hit rate, but leaves the other as the dominant bottleneck. (1) EMB optimizations, including sharding~\cite{lian2022persia,zhao2023recd}, compression~\cite{gupta2020architectural,feng2024accelerating}, prefetching~\cite{ren2025machine,chen2025toward}, and quantization~\cite{Lim2024Near_Memory} can reduce EMB access overhead but assume EMBs are the sole memory bottleneck. (2) KV cache optimizations for GR, such as cross-user KV sharing~\cite{sun2025xgr,sun2026bat,zhao2026rcllm} and KV compression~\cite{kim2025characterization}, improve cache reuse but treat KV memory in isolation. Comparisons of \sysname{} with independent KV-optimized and EMB-optimized baselines are presented in \S\ref{sec_evaluation}.

% request 调度方法
\textbf{Scheduling Optimizations for GR Inference.}
Existing GR scheduling optimizations 
such as Mooncake~\cite{qin2025mooncake}, xGR~\cite{sun2025xgr},
BAT~\cite{sun2026bat}, Flame~\cite{Guo2025FLAME}, and
ReplayGR~\cite{wang2026relaygr} focus on KV cache locality and node load under static EMB allocation. Once the EMB--KV allocation becomes dynamic, EMB transfer costs become heterogeneous 
across nodes, breaking their scheduling assumptions. Moreover, they target orthogonal objectives and do not jointly manage the EMB--KV HBM allocation. 

% 自适应资源分配说明 RL 做系统参数自适应的先例 然后指出没有人把它用在 KV-EMB 联合分区上
%得说mps
\textbf{RL for Resource Allocation in Serving Systems.} 
(1) GPU memory allocation~\cite{Oh2020Job, Saroliya2023Hierarchical} uses RL to improve utilization across co-located jobs (e.g., MPS~\cite{nvidia_mps}), focusing on inter-job scheduling. 
(2) Multi-resource allocation~\cite{Maity2024Harnessing, Udkar2025Dynamic, Wei2023Multi-Dimensional, sun2025fleetio} targets cluster-level scheduling across heterogeneous resources. 
(3) Online adaptation~\cite{chen2021drlpart, Lin2024Decentralized} addresses workload drift in datacenter partitioning. 
In contrast, we study fine-grained, intra-device contention between KV cache and EMBs within a unified HBM pool under latency-critical serving, which is not captured by prior RL formulations.
\section{CONCLUSION}
\label{sec:conclusion}
GR inference suffers from inherent HBM contention between EMB and KV caches, which prior systems fail to resolve due to isolated optimization. \sysname{} tackles this problem with a three-layer PPO-based controller that jointly reallocates HBM at runtime in an adaptive manner, together with an EMB-aware scheduler that maximizes cross-node EMB–KV reuse during serving. Experiments on three production-scale datasets over a 32-node A100 cluster show that \sysname{} reduces P99 latency by 24–38\% compared to the best static baseline, maintains $\geq$93.5\% SLO satisfaction across all workload regimes, and introduces only $32\,\mu\text{s}$ of decision overhead. These results demonstrate that joint EMB–KV management is both necessary and practical for production GR serving.

% GR inference suffers from inherent HBM contention between EMB and KV caches, which prior systems fail to resolve due to isolated optimization. \sysname{} addresses this with a three-layer PPO-based controller for adaptive HBM reallocation and an EMB-aware scheduler for cross-node cache reuse. Experiments on three production-scale datasets over a 32-node A100 cluster show that \sysname{} reduces P99 latency by 24–38\%, maintains $\geq$93.5\% SLO satisfaction, and introduces only $32\,\mu\text{s}$ of decision overhead. We hope \sysname{} motivates broader adoption of joint cache management as a design principle in future online production GR serving.
% \vspace{1.5ex}
% \noindent
% {\large\textbf{Acknowledgment}} % ← 仅在大括号内生效
% \vspace{0.5ex}
\section*{Acknowledgements}
This work is supported by the Hong Kong RGC GRF grant (No. 12201026), and Guangdong and Hong Kong Universities ``1+1+1'' Joint Research Collaboration Scheme (Project No. 2025A0505000012). We employed ChatGPT to refine the language of the entire manuscript. All generative outputs were manually reviewed and edited for technical accuracy.

\bibliographystyle{IEEEtranS}
\bibliography{refs}

% \appendices
% \input{appendix}
% \input{tech-appendix}

\end{document}